\documentclass[12pt]{article}
\usepackage{amssymb}
\usepackage{amsmath}
\usepackage{amstext}
\usepackage{graphicx,epsfig}
\usepackage{epsfig}
\usepackage{verbatim} 
\usepackage{fancyhdr}
\usepackage{fancybox}
\usepackage{color}
\usepackage{ulem,bbold}
\usepackage{enumitem}
\usepackage{subfigure}
\usepackage{bbm}
\usepackage{parskip}
\usepackage{cite}

\newcommand{\Comment}[1]{{}}
\definecolor{MyDarkBlue}{rgb}{0.60,0.40,0.80}
\usepackage[linktocpage=true]{hyperref}
\hypersetup{
colorlinks=true,
citecolor=MyDarkBlue,
linkcolor=MyDarkBlue,
urlcolor=MyDarkBlue,
}

\newcommand{\be}{\begin{equation}}
\newcommand{\ee}{\end{equation}}
\newcommand{\bea}{\begin{eqnarray}}
\newcommand{\eea}{\end{eqnarray}}
\newcommand{\beas}{\begin{eqnarray*}}
\newcommand{\eeas}{\end{eqnarray*}}
\newcommand{\nn}{\nonumber}

\newcommand{\half}{\frac{1}{2}}

\numberwithin{equation}{section}

\newcommand*\dualarrow{\ \raisebox{-1.5pt}{\epsfig{file=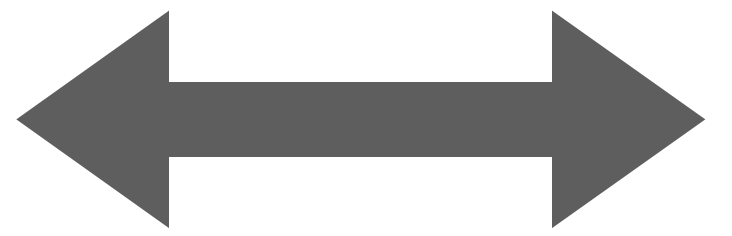,height=0.12in,width=0.25in}}\ }
\newcommand*\dualarrowline{\ \raisebox{-1.5pt}{\epsfig{file=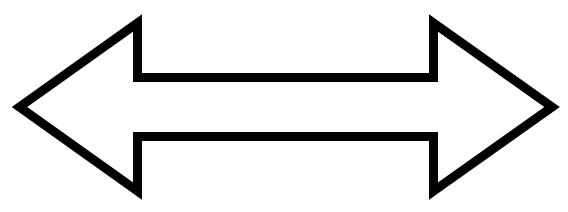,height=0.12in,width=0.25in}}\ }
\newcommand*\dualarrowcurved{\ \raisebox{-0.0pt}{\epsfig{file=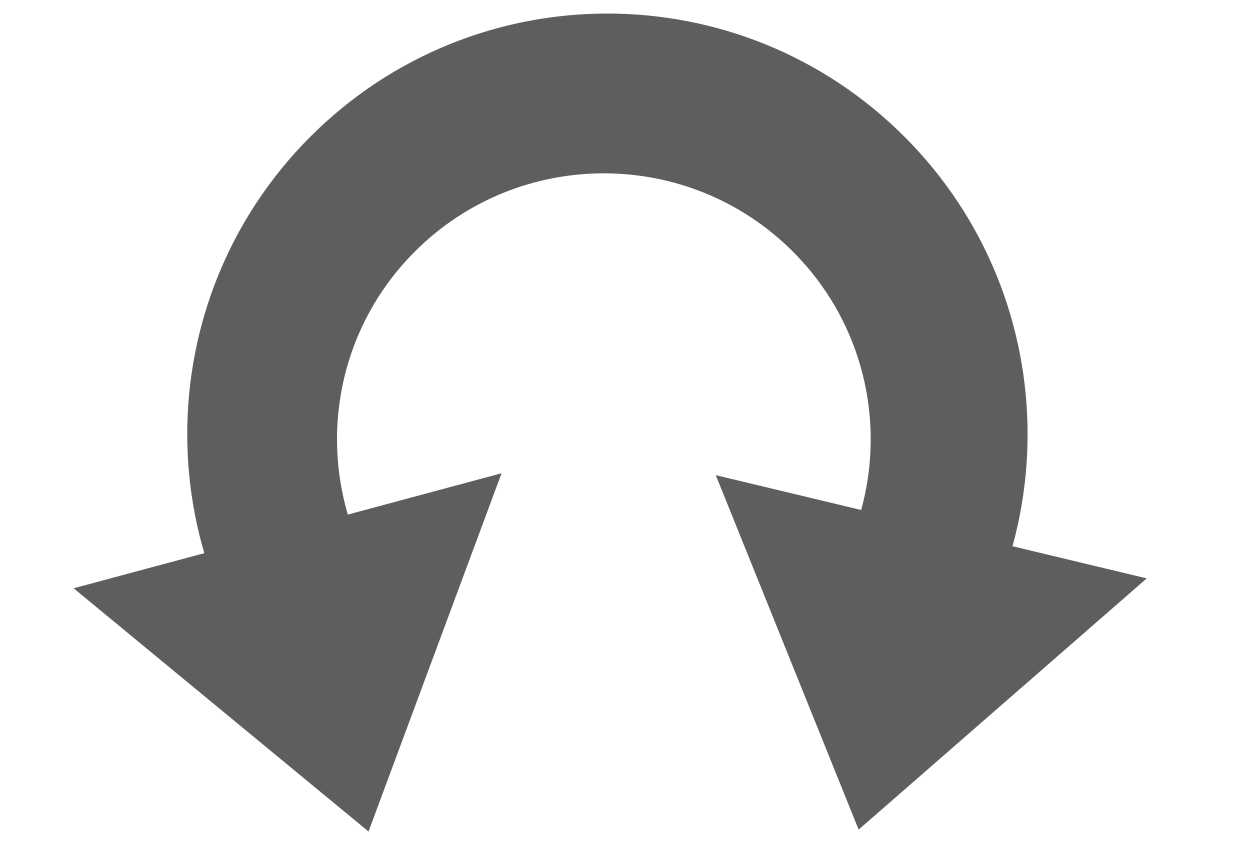,height=0.18in,width=0.26in}}\ }

\usepackage{genyoungtabtikz}
\Yboxdim{13pt} 
\Ylinethick{.8pt}

\begin{document}


\begin{center}
{\Large \bf{Dualities Among Massive, Partially Massless \\ }}
\vspace{.2cm}
{\Large \bf{and Shift Symmetric Fields on (A)dS}}
\end{center} 
 \vspace{1truecm}
\thispagestyle{empty} \centerline{
{\large {Kurt Hinterbichler}}$^{}$\footnote{E-mail: \Comment{\href{mailto:kurt.hinterbichler@case.edu}}{\tt kurt.hinterbichler@case.edu}} 
}

\vspace{1cm}

\centerline{{\it ${}^{}$CERCA, Department of Physics,}}
\centerline{{\it Case Western Reserve University, 10900 Euclid Ave, Cleveland, OH 44106}} 

 \vspace{1cm}

\begin{abstract} 

We catalog all the electromagnetic-like dualities that exist between free dynamical bosonic fields of arbitrary symmetry type and mass on (anti-) de Sitter space in all dimensions, including dualities among the partially massless and shift symmetric fields.  This generalizes to all these  field types the well known fact that a massless $p$-form is dual to a massless $(D-p-2)$-form in $D$ spacetime dimensions.  In the process, we describe the structure of the Weyl modules (the spaces of local operators linear in the fields and their derivative relations) for all the massive, partially massless and shift symmetric fields.

\end{abstract}

\newpage

\thispagestyle{empty}
\tableofcontents

\setcounter{page}{1}
\setcounter{footnote}{0}

\parskip=5pt
\normalsize

\section{Introduction}

Electromagnetic duality is a symmetry of free Maxwell electromagnetism in four dimensional flat spacetime, the appreciation of which originated with the work of Dirac \cite{Dirac:1931kp,Dirac:1948um}.  The analogous statement for massless $p$-forms is that in $D$ spacetime dimensions, a massless $p$-form is equivalent to a massless $(D-p-2)$-form.  The duality of Maxwell electromagnetism is the special case of self-equivalence when $D=4$.  Similar dualities also hold for linearized gravity and massless higher spins \cite{Hull:2001iu,Bekaert:2003az,Boulanger:2003vs,Deser:2004xt,Henneaux:2004jw,Julia:2005wg,Bunster:2012km}, for massless mixed symmetry fields \cite{Bekaert:2002dt,deMedeiros:2002qpr,Boulanger:2012mq}, and massive fields of various types \cite{Curtright:1980yk,Curtright:1980yj,Townsend:1981nu,Cecotti:1987qr,Townsend:1983xs,Casini:2002jm,Zinoviev:2005zj,Buchbinder:2008jf,Buchbinder:2009pa,Dalmazi:2011df,Morand:2012vx,Khoudeir:2014lia,Dalmazi:2020exf,Kuzenko:2020zad,Barbosa:2022zfm}.

On flat space where topological concerns are unimportant, these dualities boil down to the statement that different canonical tensor fields can carry the same propagating degrees of freedom viewed as representations of the Poincare group, or in the case of self-duality, that the  representation becomes reducible and splits into chiral halves.  For any tensor representation of the Lorentz group given by some Young tableau, there is a canonical 2-derivative action (the Fronsdal action \cite{Fronsdal:1978rb} in the case of massless spin $s$, the Labastida action for massless mixed symmetry \cite{Labastida:1987kw}, and massive versions which can be obtained by dimensional reduction) that propagates degrees of freedom in an irreducible representation of the Poincare group with little group representation given by the same tableau.  When two different tableaux are equivalent as little group representations, we have a duality (or a self-duality if a given little group representation is reducible into chiral halves).  We review these equivalences in appendix \ref{sodrepappendix} and the resulting flat space dualities in appendix \ref{flatspacedappendix}.  

On de Sitter (dS) space and anti-de Sitter space (AdS), these dualities continue to hold for $p$-forms and for the more common field types such as linearized gravity and higher spins \cite{Matveev:2004ac,Zinoviev:2005qp,Julia:2005ze,Leigh:2007wf,Basile:2015jjd,Boulanger:2018shp,Boulanger:2018adg}.  But on (A)dS there are now more interesting possibilities for irreducible representations with finite numbers of propagating degrees of freedom beyond the massive/massless binary of flat space: particles can be partially massless (PM) \cite{Deser:1983tm,Deser:1983mm} or have extended shift symmetries \cite{Bonifacio:2018zex}.   Electromagnetic-like dualities also occur for these more exotic fields on (A)dS; for example, like electromagnetism, the partially massless spin $s$ fields are known to be self-dual in $D=4$ \cite{Deser:2013xb,Hinterbichler:2014xga,Hinterbichler:2016fgl}. 

In this paper, we collect the known (A)dS dualities and add new ones involving the partially massless and shift symmetric fields, completing the list of all the dualities among all types of canonical (A)dS fields in all dimensions with propagating degrees of freedom that are finite in number (thus not covering e.g. topological theories with no propagating degrees of freedom and theories with infinite numbers of propagating degrees of freedom such as continuous spins \cite{Bekaert:2017khg}).

At the level of the action, duality invariances and equivalences in massless or PM fields are not straightforward to see covariantly, since their realization on the fundamental gauge fields is non-local \cite{Deser:1976iy,Deser:1981fr,Deser:2013xb}.  The same issues make the construction of actions for chiral fields more difficult  (see \cite{Zwanziger:1970hk,Siegel:1983es,Pasti:1995ii,Pasti:1995tn,Pasti:1996vs,Bandos:1997ui,Sen:2015nph,Sen:2019qit,Mkrtchyan:2019opf,Bansal:2021bis,Avetisyan:2022zza,Evnin:2022kqn,Evnin:2023cdf,Hull:2023dgp} for various approaches).  For massive theories, duality in the action can usually straightforwardly be seen through some kind of parent formulation involving adding in and integrating out auxiliary fields. But for massive particles of arbitrary symmetry type this rapidly becomes cumbersome, since the massive actions are generally unwieldy and require the inclusion of auxiliary fields of varying complexity (see \cite{Zinoviev:2001dt,Zinoviev:2002ye,Zinoviev:2003dd,Hallowell:2005np,Buchbinder:2005ua,Buchbinder:2006ge,Buchbinder:2007ix,Buchbinder:2008ss,Zinoviev:2008ze,Zinoviev:2016mxh} for various approaches).  Even at the level of the equations of motion, duality invariance beyond the $p$-forms can be quite subtle, and appropriate off-shell or partially off-shell formulations must be used.

However, at the level of the gauge invariant local operators, i.e. the local on-shell observables of the theory, duality is straightforward: there is an invertible map involving the epsilon tensor that takes the local operators of one theory into the local operators of the dual theory (or the same theory, in the case of self-duality).  For example, in pure electromagnetism in $D=4$, the basic gauge invariant operator is the Maxwell field strength $F_{\mu\nu}$, and the map $F_{\mu\nu}\rightarrow \tilde F_{\mu\nu}\equiv {1\over 2} \epsilon_{\mu\nu\rho\sigma}F^{\rho\sigma}$ implements the duality.

The local operators are all products of powers of the field and derivatives, modulo the equations of motion.  In the massless, PM and shift symmetric cases, they must also be invariant under the relevant gauge transformations or shift symmetries.  The building blocks of these local operators are those that contain a single power of the field, with arbitrary powers of the derivatives.  The set of such single-field operators forms an algebraic structure known as the Weyl module.  One can think of the Weyl module as a set of states, each of which represents a distinct operator.  Each state satisfies the Klein-Gordon equation with some mass that we can think of as an energy level.  Acting with derivatives, we can move between these states as we do with raising/lowering operators in quantum mechanics.  

When two fields are dual, their Weyl modules are equivalent once all the operators are brought to a minimal form using the epsilon tensor.  In this way, we will be able to see and understand all the dualities that occur among the various (A)dS fields.  In what follows, in Section \ref{adssec2e} we review the various types of fields, explain the structure of their Weyl modules, and show how to see the dualities.  We then tabulate all the dualities in Section \ref{adstabesecees}.

\textbf{Conventions:}   We work in $D$ dimensional Lorentzian spacetimes using the mostly plus metric signature and the curvature conventions of \cite{Carroll:2004st}.   Formulas are expressed in terms of dS$_D$, where we use $H$ to denote the dS Hubble scale, so that the Ricci scalar is $R={D(D-1)H^2}>0$.  All the results also apply to AdS$_D$ upon making the replacement $H^2\rightarrow -1/L^2$ with $L$ the AdS scale.
Tensors are symmetrized and anti-symmetrized using unit weight ({\it e.g.} $t_{[\mu\nu]}=\half \left(t_{\mu\nu}-t_{\nu\mu}\right)$) and we use $(\cdots)_T$ to denote the fully traceless symmetric part of the enclosed indices.  Young tableaux are written as $[s_1,s_2,\ldots,s_p]$, where $s_i$ denotes the number of squares in the $i$-th row, $s_1\geq s_2\geq \ldots \geq s_p$.
We also use the shorthand notation of using an exponent to denote multiple rows with the same length, e.g. $[2,1^3]\equiv [2,1,1,1]$.  Tableaux are always taken to be fully traceless, and we use the manifestly anti-symmetric convention.  The projector onto the symmetries of a tableau is denoted ${\cal Y}^T_{[s_1,\ldots,s_p]}$.  The indices to be projected are either clear from context or explicitly shown in the subscript of the projector.

\section{Fields, Weyl modules and dualities\label{adssec2e}}

We start by reviewing the massive, PM and shift-symmetric fields on (A)dS in Section \ref{massivesec}, and their Weyl modules, or spaces of local linear operators, in Section \ref{weylmodsec} (from now on, unless otherwise stated, we take the term ``partially massless'' to also include what are usually called the massless cases).  We then use them in Section \ref{secweylmoddualsec} to illustrate the dualities.

\subsection{Massive, PM and shift-symmetric fields\label{massivesec}}

\textbf{Massive fields:} A massive field on (A)dS$_D$ is carried by a tensor with the index symmetries of a $p$ row Young tableau labelled by its row lengths, $[s_1,\ldots,s_p]$.  On-shell, the field is completely traceless, divergenceless in all indices, and vanishes when acted on by the  Klein-Gordon operator, 
\be  \nabla^2-\tilde m^2_{[s_1,\ldots,s_p]}\,,\label{baremassiniee}\ee
with a mass $\tilde m^2_{[s_1,\ldots,s_p]}$ that we call the {\it bare mass} of the field\footnote{For the spin-$s$ fields ${[s]}$, the mass $m_{[s]}^2$ is usually used, which is related to the bare mass by 
\be \tilde m^2_{[s]}=m_{[s]}^2 +\left[s+D-2-(s-1)(s+D-4)\right]H^2\, .\ee
It is shifted so that $m_{[s]}^2=0$ is the massless value.  For the $p$-form fields ${[1^p]}$, the mass $m_{[1^p]}^2$ is usually used, which is related to the bare mass by
\be \tilde m^2_{[1^p]}=m_{[1^p]}^2 +p(D-p)H^2\, .\ee
It is shifted so that $m_{[1^p]}^2=0$ is the massless value. 
}.  
The dual conformal field theory (CFT) conformal dimensions $\Delta_\pm$ of the field are found from the bare mass in \eqref{baremassiniee} by solving for the greater and lesser roots of the equation
\be {\tilde m^2_{[s_1,\ldots,s_p]}/ H^2}=\Delta(d-\Delta) +\sum_{i=1}^p s_i \, , \label{mixsymmassde2ree}\ee
where $d\equiv D-1$ is the dual CFT dimension.

\textbf{PM fields:}  At particular values of the mass, the field acquires enhanced gauge symmetries and loses some of its propagating degrees of freedom.   These are the PM points, and the classification of points for which this happens is detailed in \cite{Metsaev:1995re,Metsaev:1997nj,Alkalaev:2003qv,Boulanger:2008up,Boulanger:2008kw,Skvortsov:2009zu,Skvortsov:2009nv,Basile:2016aen}.  A brief summary is as follows:  The PM points occur when some squares of the tableau in the last row of a block (i.e. from a row which is longer than the row right below it) are `activated'.   If it is squares from the $q$-th row being activated, then the number of activated squares can range from $t=1,2,\ldots, s_q-s_{q+1}$.  The depth parameter $t$ is used to indicate how many squares are activated\footnote{The paper \cite{Hinterbichler:2022vcc} stubbornly used a different convention for $t$, $t_{\rm there}=s_q-s_{q+1}-t_{\rm here}$, which annoyingly did not conform to the rest of the literature.}.  Removing the squares that are activated, we get the Young tableau of the gauge parameter under which the PM field transforms, with each activated square becoming a derivative in the gauge transformation rule.  

We will use the following shorthand notation to designate the PM fields: we put a $\nabla^t$ underneath the $q$-th row to indicate that $t$ squares of the $q$-th row are activated,
\be \underset{\ \ \nabla^t}{[s_1,\ldots, s_q,\ldots s_p]}\,.\ee
For example, the massless 2-form is $\underset{\ \ \  \nabla}{[1,1]}$, the massless graviton is $\underset{\nabla}{[2]}$, and the PM graviton is $\underset{\nabla^2}{[2]}$.  

The PM points happen at the values of the dual CFT dimension given by
\be \Delta_+= d - q + s_{q} - t\,.\label{PMcftdimmese}\ee
Putting this into \eqref{mixsymmassde2ree}, we get the PM masses,
\be \tilde m^2_{\underset{\nabla^t}{[s_1,\ldots, s_q,\ldots,s_p]}}/H^2=(D-1 - q + s_{q} - t)(q - s_{q} + t)+  \sum_{i=1}^p s_i \,. \ \label{gensymshsmassee}\ee

The basic gauge invariant field strength tensor for a PM field, generalizing the Maxwell field strength and the linear Weyl tensor for massless spin 1 and spin 2 respectively, can be found by making a new tableau by adding $s_q-s_{q+1}-t+1$ 
new boxes to the row below the activated row, putting the PM field into this new tableau while filling the new boxes with derivatives, then projecting onto the new tableau, 
\be  \underset{\ \ \nabla^t}{[s_1,\ldots, s_q,\ldots s_p]}\ {\rm field\ strength:}\ \   [s_1,\ldots,s_{q},s_{q}-t+1,s_{q+2},\ldots,s_p] \, . \label{PMieldstrengthbe} \ee
By virtue of its definition in terms of derivatives of the field, this field strength satisfies a Bianchi identity given by adding $s_{q+1}-s_{q+2}+1$  boxes corresponding to derivatives on the $(q+2)$-th row, giving a tableau,
\be  \underset{\ \ \nabla^t}{[s_1,\ldots, s_q,\ldots s_p]}\ {\rm Bianchi\ identity:}\ \   [s_1,\ldots,s_{q},s_{q}-t+1,s_{q+1}+1,s_{q+3}\ldots,s_p]\,.\label{PMieldstrengthbiae}\ee

\textbf{Shift symmetric fields:}  At other particular values of the mass, the massive field acquires enhanced shift symmetries, generalizing the shift symmetries of the massless scalar and galileon \cite{Nicolis:2008in}.   The classification of shift symmetries is detailed in \cite{Bonifacio:2018zex} for spin $s$ and \cite{Hinterbichler:2022vcc} for general mixed symmetry.  A brief summary is as follows:  the shift symmetry points of a massive field with tableau $[s_1,\ldots,s_p]$ are labelled by an integer $k=0,1,2,\ldots$ called the {\it level}, and occur at the values of the dual CFT dimension given by
\be \Delta_+=d +k  +  s_1\,.\label{genmixwsconfdde}\ee  

We will use the following shorthand notation to designate the shift symmetric fields: we put a $k$ underneath the tableau to indicate that it is a level $k$ shift symmetric field,
\be \underset{k}{[s_1,\ldots, s_p]}\,.\ee
For example, the massless scalar is $\underset{0}{[0]}$.  From \eqref{genmixwsconfdde}, \eqref{mixsymmassde2ree}, the shift symmetric masses are ,
\be \tilde m^2_{\underset{k}{[s_1,\ldots, s_p]}}/H^2=(D-1+k+s_1)(-k-s_1)+  \sum_{i=1}^p s_i \,. \ \label{gensymshsmassee}\ee

This shift symmetric field at level $k$ can be thought of as the longitudinal mode which decouples as a massive field of type $[s_1+k+1,s_2,\ldots, s_p]$ approaches its PM point at $\Delta_+=d - 1 +  s_1$ where $t=k+1$ squares of the top row are activated.  We call this the PM field {\it associated} to the shift symmetric field.

We will consider the shift symmetries to be a kind of gauge symmetry, so that we are required to mod out by the shifts and consider only quantities that are invariant under them.  The basic field strength tensor invariant under the shift symmetry is made from adding $k+1$ squares signifying derivatives to the first row of the field's tableau, giving a tableau with the same shape as that of the associated PM gauge field \cite{Bonifacio:2018zex,Hinterbichler:2022vcc}, 
\be   \underset{k}{[s_1,\ldots s_p]}\ {\rm field\ strength:}\ \   [s_1+k+1,s_2,\ldots, s_p]\,.\label{shiftfieldstrengthe}\ee
By virtue of its definition in term of derivatives of the field, this field strength satisfies a Bianchi identity given by adding $s_1-s_2+1$ squares corresponding to derivatives on the second row of the field strength, giving a shape identical to that of the field strength of the associated PM field,
\be  \underset{k}{[s_1,\ldots s_p]}\ {\rm Bianchi\ identity:}\ \   [s_1+k+1,s_1+1,\ldots, s_p]\,.\label{shiftfieldstrengthbe}\ee

\subsection{Weyl modules and their PM and shift symmetric points\label{weylmodsec}}

In the higher-spin literature, there is a structure known as the {\it Weyl module} that encodes the primitive on-shell non-trivial gauge invariant local operators of a theory.  
The content of the Weyl module for various spins can be found in \cite{Boulanger:2008up,Boulanger:2008kw,Skvortsov:2009zu,Skvortsov:2009nv,Alkalaev:2009vm,Ponomarev:2010st,Alkalaev:2011zv,Boulanger:2014vya,Boulanger:2015mka,Khabarov:2019dvi} (see \cite{Lopatin:1987hz,Vasiliev:2001wa,Alkalaev:2003qv} for some earlier works).   It usually appears in the context of the unfolded frame-like formulations of higher spin field equations (see \cite{Didenko:2014dwa} for a review).  Here we will try to describe the general case in a more concrete way that will be sufficient for our purposes, will make the PM and shift symmetric points easy to see, and will make the dualities between fields clear.  We illustrate the structure starting with the simplest example of a massive scalar, and working up to the most general case.

\textbf{Scalar:}
Consider a massive scalar $\phi$ with mass $\tilde m_{[0]}$ on (A)dS$_D$, which satisfies the standard Klein-Gordon equation,
\be \left(\nabla^2 -\tilde m_{[0]}^2\right)\phi=0.\label{phieome}\ee
The local operators of the theory are all polynomials made from products of $\phi$'s and covariant derivatives acting in all possible ways, modulo the equation of motion \eqref{phieome}.  The primitive building blocks of these are those with a single $\phi$, from which the others can be assembled as products.  The primitive building blocks containing a single $\phi$ are easy to describe: at any derivative order $l$, since all derivative act on a single $\phi$, any antisymmetric part can be removed in favor of lower derivatives by commuting the covariant derivatives and using the maximal symmetry of the background.  Any trace part can similarly be removed using the equations of motion \eqref{phieome}.  Thus the only way derivatives can act on a single $\phi$ modulo \eqref{phieome} is via the symmetric traceless expressions
\be [l]\equiv  \nabla_{(\mu_1}\nabla_{\mu_2}\cdots \nabla_{\mu_l)_T}\phi,\ \ \ l=0,1,2,\ldots\, ,\label{phiweylmodu}\ee
and so these give the complete list of on-shell non-trivial local operators linear in $\phi$.   

Each tensor $[l]$ itself satisfies a Klein-Gordon equation with masses given as follows\footnote{The algebra of operators on constant curvature spaces \cite{Hallowell:2005np} is useful for finding equations such as these.}, 
\be \left(\nabla^2-\tilde m_{[0]}^2-l(D+l-2)H^2\right)[l]=0\, .\label{scalarmasemodee}\ee
If we take a single derivative of $[l]$, the result can be decomposed on shell into a symmetric traceless tensor of rank $l+1$ and a symmetric traceless tensor of rank $l-1$ (the traceless mixed symmetry part vanishes identically).  The symmetric traceless rank $l+1$ part is given by the symmetrized traceless derivative, which gives precisely the operator $[l+1]$, with unit prefactor.  The symmetric traceless rank $l-1$ part is the divergence, and gives the operator $[l-1]$ with a prefactor as follows,
\be \nabla^{\mu_l}\nabla_{(\mu_1}\cdots \nabla_{\mu_l)_T}\phi = {D-3+l\over D+2l-4}\left(\tilde m_{[0]}^2+(l-1)(l+D-2) \right)  \nabla_{(\mu_1}\cdots \nabla_{\mu_{l-1})_T}\phi\,.\label{scalargrade}\ee

We can think of the space of operators $[l]$ as analogous to a set of quantum states or energy levels, with energies given by the masses in \eqref{scalarmasemodee}, with the basic operator $\phi$ as the ground state, and with lowering and raising operators given by the divergences and symmetrized traceless gradients respectively.  This structure of states and the one-derivative relations among them is the Weyl module for a generic massive scalar, which we can illustrate as follows:
\be  [0]\ {\rm Weyl\ module:}\  \raisebox{-19pt}{\epsfig{file=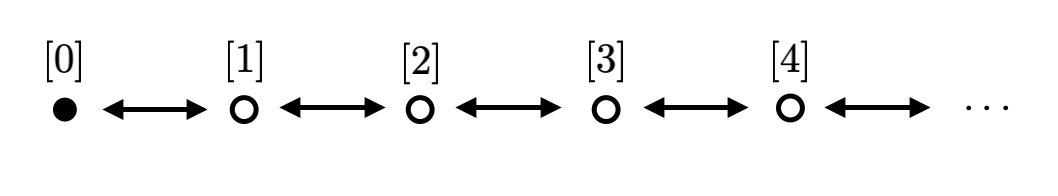,height=0.7in,width=3.6in}} \ \ \ . \label{scalarwmode}\ee 
Here the arrow directions pointing left are the divergences and the arrow directions pointing right are the symmetrized traceless gradients.  The ground state $[0]$ is denoted with a solid black dot, and the excited states above it by hollow dots.

When the scalar mass takes the mass value of \eqref{gensymshsmassee} corresponding to the $k$-th shift symmetric scalar, $m^2_{\underset{k}{[0]}}/H^2=-k(k+D-1)$, the prefactor in \eqref{scalargrade} for the state $l=k+1$ vanishes, and so we cannot reach the state $[k]$ starting from the state $[k+1]$.  This means that the arrow pointing left from the state $[k+1]$ in \eqref{scalarwmode} is broken and thus the states $[l]$ with $l\geq k+1$ form a submodule, with $[k+1]$ as the new ground state, as illustrated here:
\be \underset{k}{[0]}\ {\rm Weyl\ module:}\  \raisebox{-24pt}{\epsfig{file=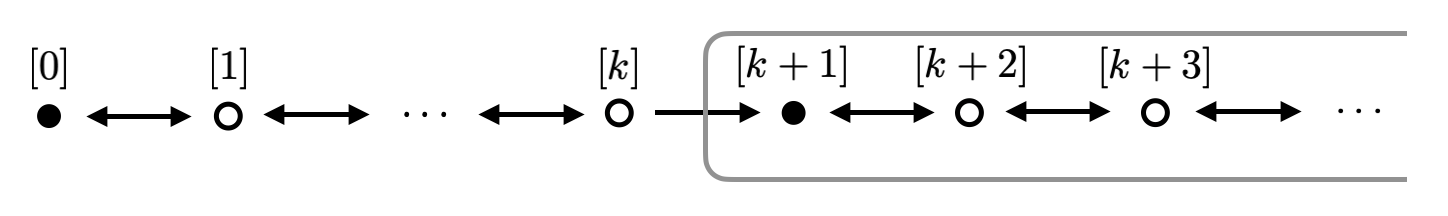,height=0.8in,width=4.9in}} \ \ \ . \label{scalarkwmode}\ee  
The submodule is enclosed in the grey boundary, and the new ground state $[k+1]$ is now indicated in black.
This new ground state satisfies the Klein-Gordon equation 
\be \left(\nabla^2 -(D-1+k)H^2\right) [k+1]=0\,.\label{sckpp1fem2e}\ee
  The new submodule it generates is the set of local operators for the level $k$ shift symmetric field, i.e those that are invariant under the shift symmetry. The new ground state $[k+1]$ is the lowest such operator, i.e. the one with the fewest number of derivatives on the field. It has $k+1$ derivatives and is nothing but the shift invariant field strength \eqref{shiftfieldstrengthe}.

Though the Weyl module explicitly displays only the on-shell non-trivial operators and the derivative relations among them, it also implicitly encodes the equations of motion that these operators satisfy (on top of the Klein-Gordon equation).  Any derivative that does not correspond to an arrow that is displayed in the Weyl module leads to a vanishing tensor and thus an equation of motion.  For example, consider the ground state operator $[k+1]$ for the level $k$ shift field, and consider all its possible first derivatives.  These are classified by the tensor product $[1]\otimes [k+1]=[k+2]\oplus [k+1,1] \oplus [k]$. $[k]$ is the divergence, and there is no arrow for this in the Weyl module \eqref{scalarkwmode}, so one of the equations of motion is that this operator is divergenceless.  $[k+1,1]$ is the traceless mixed symmetry derivative; a tensor with this symmetry does not appear in the Weyl module, which tells us that this expression vanishes identically, so this is a Bianchi identity-like equation on $[k+1]$.  $[k+2]$ is the traceless symmetrized derivative, which does appear as the first excited state in the module, so this is not an equation of motion.  Thus $[k+1]\equiv F_{\mu_1\cdots \mu_{k+1}}$ satisfies two one-derivative equations of motion analogous to the Maxwell equations,
\be \nabla^{\mu_{k+1}}F_{\mu_1\cdots \mu_{k+1}}=0,\ \ \ \nabla_{[\mu_{k+2}}F_{\mu_1]\mu_2\cdots \mu_{k+1}}-{\rm traces}=0.\label{shsbasiee}\ee
Note that the second of these is precisely the Bianchi identity \eqref{shiftfieldstrengthbe}. 
Proceeding to higher derivatives will give additional higher derivative equations of motion, but these will all be derivative consequences of \eqref{shsbasiee} and the Klein-Gordon equations, so we can consider \eqref{shsbasiee} as the basic equations of motion.  

\textbf{Vector:}
Next consider a massive vector $A_\mu$ with mass $\tilde m_{[1]}$, whose on-shell equations are
\be \left(\nabla^2 - \tilde m_{[1]}^2\right)A_\mu=0,\ \ \ \nabla^\mu A_\mu=0.\ee
The space of on-shell non-trivial operators linear in $A_\mu$ now consists of a set of symmetric traceless tensors,
\be [l]\equiv \nabla_{(\mu_1}\cdots\nabla_{\mu_{l-1}}A_{\mu_l)_T}\,,\ \ \ l=1,2,3,\ldots\, , \ee
and a set of mixed symmetry tensors 
\be [l,1] \equiv  {\cal Y}_{[l,1]}^T\left( \nabla_{\mu_2}\cdots \nabla_{\mu_{l}} \nabla_{\mu_1}A_{\nu}\right) \,,\ \ \ l=1,2,3,\ldots\, . \ee
These all satisfy a Klein-Gordon equation on shell,
\be \left(\nabla^2 -\tilde m_{[1]}^2-\left(l^2+(D-2)l-2\right)H^2\right)[l,1]=0\,,\ \ \ \left(\nabla^2 - \tilde m_{[1]}^2-(l-1)(l+D-1)H^2\right)[l]=0\,.\ee
We can think of the $l=1$ states, $[1]= A_\mu$ and $[1,1]=  \nabla_{[\mu}A_{\nu]}$, as two ground states\footnote{Here the energy level analogy breaks down somewhat because we call both of these ground states despite having different masses in their Klein-Gordon equations. A similar caveat applies to many of the examples to follow.} and the higher $l$'s as excited states built off them, and arrange everything in a module as follows,
\bea  
&& [1]\ {\rm Weyl\ module:}\  \raisebox{-48pt}{\epsfig{file=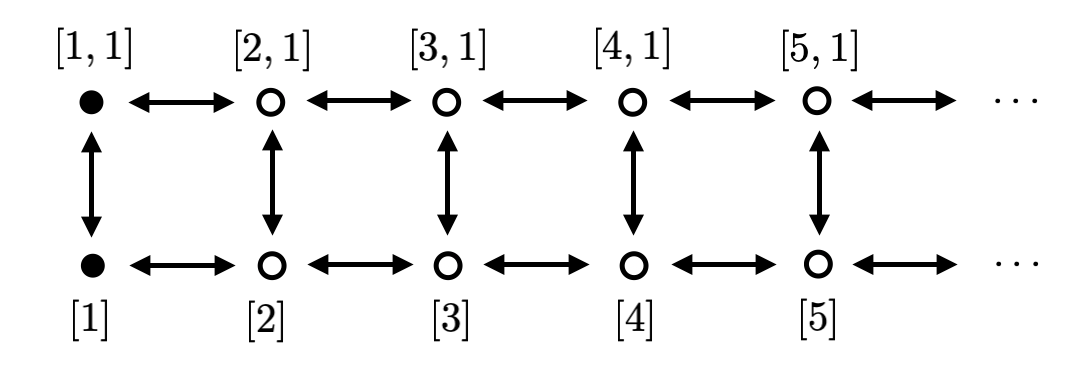,height=1.4in,width=4.2in}} \ \ \ \ . \nn\\  \label{massivevmodulee}
\eea
Here each arrow direction represents taking a derivative and then projecting onto the target tableau, so the downward and leftward arrows represent divergences, the upwards arrows are anti-symmetrized gradients and the rightward arrows are symmetrized gradients.  For example, the downward arrow direction from $[1,1]$ to $[1]$ tells us that if we take a divergence of $[1,1]$ we arrive at $[1]$,
\be \nabla^\nu \nabla_{[\mu}A_{\nu]}=-{1\over 2} \left( \tilde m_{[1]}^2-(D-1)H^2\right)A_\mu.  \ee
As another example, consider the divergences of $[2,1]$: there are two different divergences one can take of $[2,1]$, but both lead to a linear combination of $[1,1]$ and $[2]$, for example taking the divergence on the bottom square of $[2,1]$ gives
\bea && \nabla^\alpha {\cal Y}^T_{\resizebox{.5cm}{!}{\gyoung(\mu\nu ,\alpha )}}\left[ \nabla_\mu\nabla_\nu A_\alpha \right]= \nn\\ 
&& \quad {D-2\over 3(D-1)}\left(  \tilde m_{[1]}^2+(D-1)H^2 \right)\nabla_{[\mu}A_{\nu]}-{D-2\over 3(D-1)}\left(  \tilde m_{[1]}^2-(D-1)H^2 \right) \nabla_{(\mu}A_{\nu)_T}\,. \nn\\
\eea
The arrow from $[2,1]$ to $[1,1]$ represents the first term on the bottom line and the arrow from $[2,1]$ to $[2]$ represents the second.

When the vector mass takes one of the shift symmetric mass values \eqref{gensymshsmassee}, $\tilde m_{\underset{k}{[1]}}^2/H^2=-(k+1)(k+D)+1$, $k=0,1,2,\ldots$, the leftward arrows leading from the $l=k+2$ states to the $l=k+1$ states vanish, and the states with $l\geq k+2$ split off into an irreducible module with the new ground states $[k+2,1]$, $[k+2]$, as illustrated here:
\bea  
&& \underset{k}{[1]}\ {\rm Weyl\ module:}\  \raisebox{-42pt}{\epsfig{file=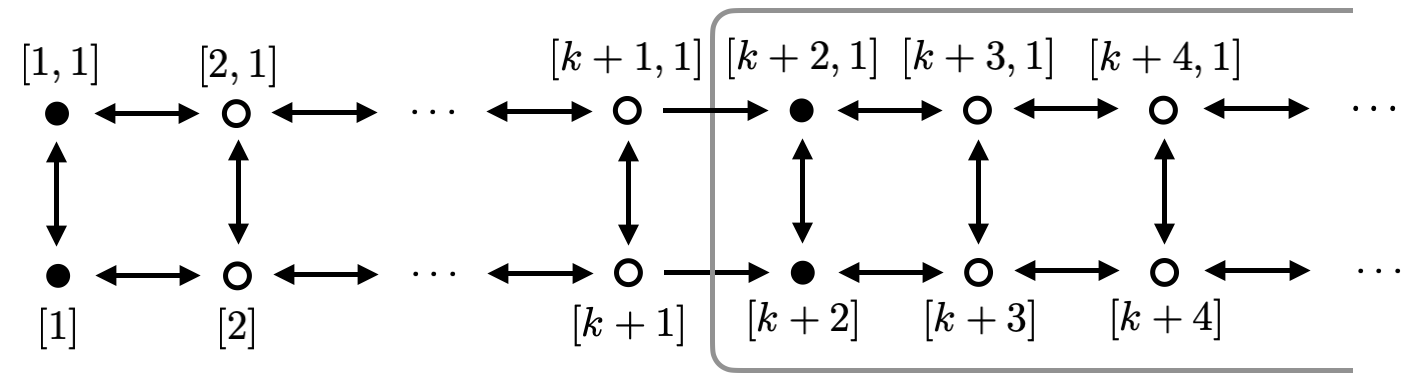,height=1.2in,width=4.2in}}\ \ \ \ . \nn \\  \label{massivevkmodulee}
\eea
This is the level $k$ shift symmetric vector, and the states in the new module, circled in grey, are the ones invariant under the generalized shift symmetry of level $k$.  Of these, the one with the fewest number of derivatives is the state $[k+2]$, and this is the shift invariant field strength \eqref{shiftfieldstrengthe}.

When the vector bare mass takes the mass value $\tilde m_{\underset{\nabla}{[1]}}^2=(D-1)H^2$, the vector becomes massless, and gets the on-shell gauge invariance $\delta A_\mu=\nabla_\mu \Lambda$, where $\Lambda$ is the on-shell scalar gauge parameter, which satisfies the equations of a $\underset{0}{[0]}$ field, namely $\nabla^2  \Lambda=0$.  At this value of the mass, all the downwards arrows from the top row of \eqref{massivevmodulee} to the bottom row vanish, so the top row separates off into its own irreducible module as illustrated here:
\bea  
&& \underset{\nabla}{[1]}\ {\rm Weyl\ module:}\  \raisebox{-40pt}{\epsfig{file=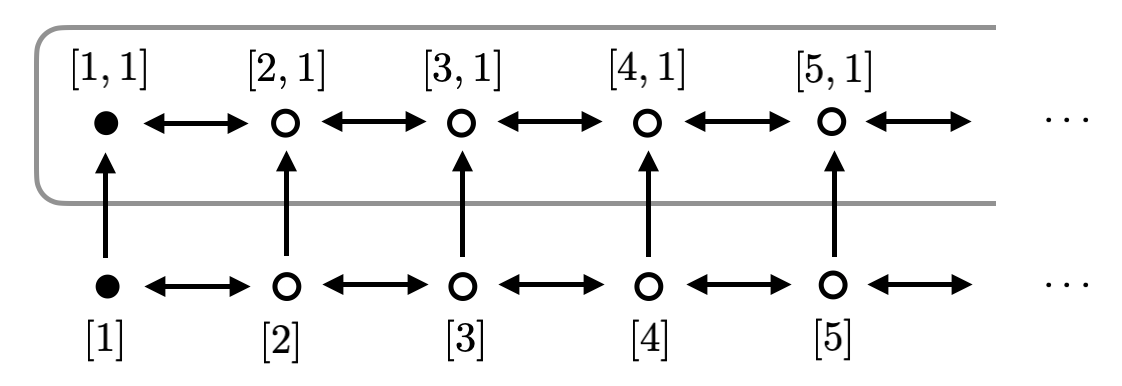,height=1.2in,width=3.5in}}\ \ \ \ . \nn\\ 
 \label{massivevmoduleme2}
\eea
This new module, comprised of the states circled in grey, is the module of gauge invariant operators.  It is generated from the ground state $[1,1]\propto F_{\mu\nu}\equiv \nabla_\mu A_\nu -\nabla_\nu A_\mu$, which is nothing but the Maxwell field strength, i.e. the ground state is the field strength \eqref{PMieldstrengthbe} for $\underset{\nabla}{[1]}$.  

Equations of motion are encoded by the absence of states and arrows in the module; for example, for the massless vector, the possible first order derivatives of the field strength are $[1]\otimes [1,1]=[2,1]\oplus [1,1,1]\oplus [1]$, and the two which do not appear in \eqref{massivevmoduleme2} are $[1]$ and $[1,1,1]$, corresponding to the divergence and anti-symmetric gradient, respectively.  The vanishing of these are the free Maxwell equations.  
\be \nabla^\nu F_{\mu\nu}=0\, ,\ \ \ \nabla_{[\mu}F_{\nu\rho]}=0\,.\ee
The second of these, the $[1,1,1]$ equation, is the Bianchi identity \eqref{PMieldstrengthbiae} for $ \underset{\nabla}{[1]}$.  

Note that the states that are removed from the Weyl module at the massless point, those in the bottom row of \eqref{massivevmoduleme2}, are precisely those of a $k=0$ shift symmetric scalar, $\underset{0}{[0]}$, which is the longitudinal mode of the massive spin-1 that splits off as it becomes massless; this is how the Weyl module encodes the St\"uckelberg mechanism.

\textbf{spin $s$:}
For a massive spin-$s$ field $\phi_{\mu_1\cdots \mu_s}$, the ground states of the module consist of the tensors $[s],\ [s,1],\ [s,2],\ \ldots\ , [s,s]$, where $[s]=\phi_{\mu_1\cdots \mu_s}$ is the original field, $[s,1]={\cal Y}^T_{[s,1]}\left[\nabla_{\nu_1}\phi_{\mu_1\cdots \mu_s}\right]$, and so on up to $[s,s]={\cal Y}^T_{[s,s]}\left[\nabla_{\nu_1}\cdots \nabla_{\nu_s} \phi_{\mu_1\cdots \mu_s}\right]\equiv W_{\mu_1\cdots\mu_s,\nu_1\cdots\nu_s}$, which is the generalized Weyl tensor \cite{deWit:1979sib,Damour:1987vm} for a massless spin $s$ field.  On top of these ground states, we build out excited states by acting with more symmetrized derivatives, giving the full set of states $[l],\ [l,1], \ldots,\ [l,s]$ for $l\geq s$, with $l=s$ the ground states.  There will be arrows corresponding to taking various derivatives and then Young symmetrizing, which fill out a module as follows,
\bea && [s]\ {\rm Weyl\ module:}\ \ \  \raisebox{-100pt}{\epsfig{file=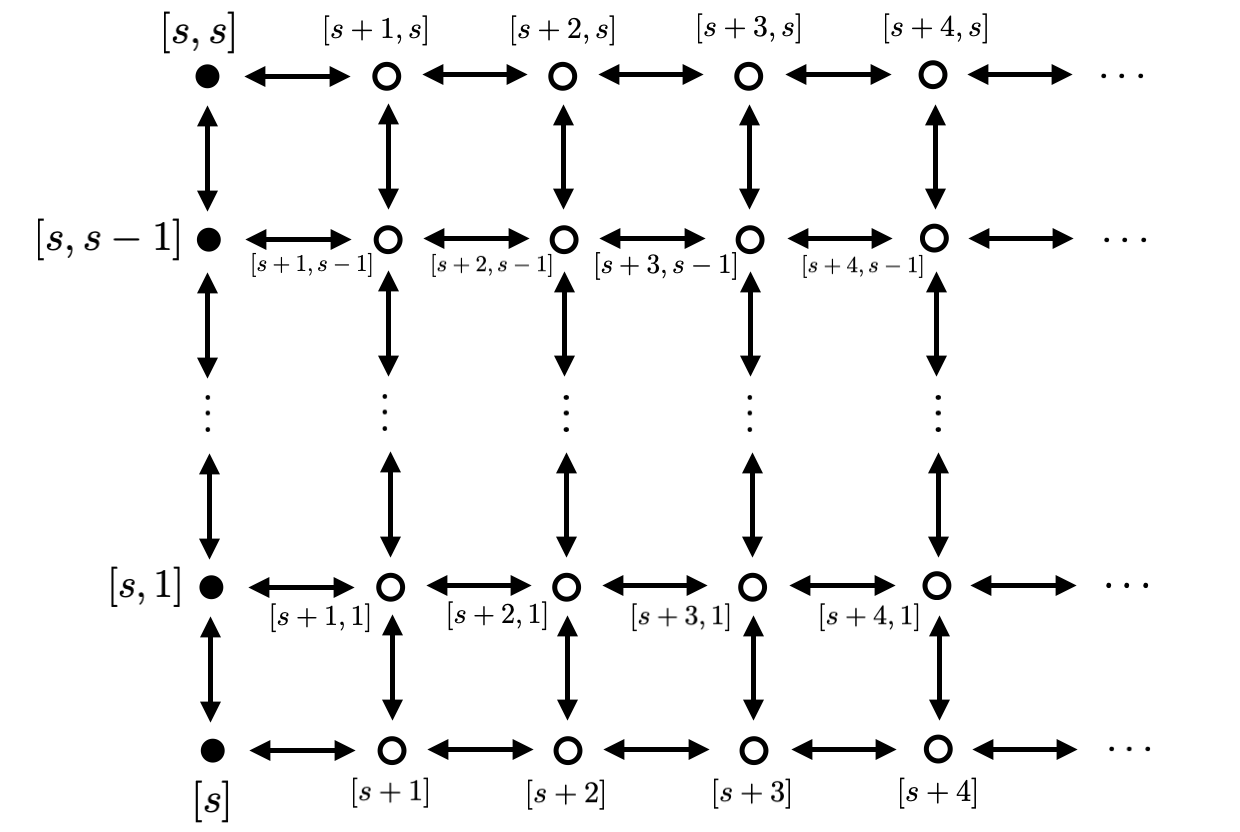,height=2.9in,width=4.1in}} \ \ \ \ . \nn\\ 
\label{massivevmoduleespins2}
\eea

When the mass takes the level $k$ shift symmetric value, $\tilde m^2_{\underset{k}{[s]}}/H^2=- (k + s) (k + s+D-1)+s$, the leftward arrows from the $l=k+s+1$ to $l=k+s$ column of \eqref{massivevmoduleespins2} vanish, isolating the states with $l\geq k+s+1$ into an irreducible module.  These are the operators invariant under the level $k$ generalized shift symmetry.  The basic field strength \eqref{shiftfieldstrengthe} is the $[k+s+1]$ ground state, which is the state within the new module with the fewest number of derivatives.

At the massless value of the mass, $\tilde m^2_{\underset{\nabla}{[s]}}/H^2= - (s-2) ( s+D-3)+s$, the downwards arrows from the top row of \eqref{massivevmoduleespins2} vanish, isolating the top row into an irreducible module, which are the operators invariant under the on-shell massless gauge symmetry $\delta \phi_{\mu_1\cdots \mu_s}=\nabla_{(\mu_1}\Lambda _{\mu_2\cdots \mu_{s})_T}$, where $\Lambda _{\mu_1\cdots \mu_{s-1}}$ is an on-shell $\underset{0}{[s-1]}$ gauge parameter.  These are generated by the generalized Weyl tensor $[s,s]$, which is the field strength \eqref{PMieldstrengthbe} for $\underset{\nabla}{[s]}$ and is the new ground state of the module.  The absence of arrows to $[s,s-1]$ and to $[s,s,1]$ shows that the generalized Weyl tensor satisfies generalized Maxwell equations,
\be \nabla^{\mu_1}W_{\mu_1\cdots\mu_s,\nu_1\cdots\nu_s}=0\, ,\ \ \  \nabla_{[\rho}W_{\mu_1|\cdots\mu_s|,\nu_1]\cdots\nu_s}-{\rm traces}=0\,.\ee  
The second of these, $[s,s,1]$, is the Bianchi identity \eqref{PMieldstrengthbiae} for $ \underset{\nabla}{[s]}$.  The states in the massive module that are not part of the massless module, all but the top row of \eqref{massivevmoduleespins2}, are precisely those of the Weyl module for a level $k=0$ shift symmetric spin $s-1$ field, $\underset{0}{[s-1]}$, which is the longitudinal mode that decouples in the massless limit.

At the depth $t$ partially massless value, $\tilde m_{\underset{\nabla^t}{[s]}}^2/H^2=- ( s - t-1) (s - t+D-2)+s$, the downwards arrows from the $t$-th row of \eqref{massivevmoduleespins2} vanish, isolating the top $t$ rows into an irreducible module, which are the operators invariant under the on-shell PM gauge symmetry $\delta \phi_{\mu_1\cdots \mu_s}=\nabla_{(\mu_1}\cdots \nabla_{\mu_t}\ \Lambda _{\mu_{t+1}\cdots \mu_{s})_T}$, where $\Lambda _{\mu_1\cdots \mu_{s-t}}$ is an on-shell $\underset{t-1}{[s-t]}$ gauge parameter.  These are generated by the generalized Weyl tensor $[s,s-t+1]$, which is the field strength \eqref{PMieldstrengthbe} for $\underset{\nabla^t}{[s]}$ and is the new ground state of the module.  The absence of arrows to $[s,s-t]$ and $[s,s-t+1,1]$ shows that it satisfies generalized Maxwell equations 
\be \nabla^{\mu_1}W_{\mu_1\cdots\mu_s,\nu_1\cdots\nu_{s-t+1}}=0\, ,\ \ \  \nabla_{[\rho}W_{\mu_1|\cdots\mu_s|,\nu_1]\cdots\nu_{s-t+1}}=0\, .\ee   
The second of these, $[s,s-t+1,1]$ is the Bianchi identity \eqref{PMieldstrengthbiae} for $\underset{\nabla^t}{[s]}$.    The states in the massive module that are not part of the depth $t$ PM module, all but the top $t$ rows of \eqref{massivevmoduleespins2}, are precisely those of the Weyl module for a level $k=t-1$ shift symmetric spin $s-t$ field, $\underset{t-1}{[s-t]}$, which is the longitudinal mode that decouples in the massless limit.

\textbf{General case:}
Now consider a massive field in the general tableau $[s_1,\ldots,s_p]$.  The procedure for generating the Weyl module is as follows.  First take the tableau and add another row of length $s_1$ to the top, to create the tableau $[s_1,s_1,\ldots,s_p]$.  Then branch this down one dimension using the tableau branching rule
\be [u_1,\ldots,u_p]\rightarrow \bigoplus_{v_1=u_2}^{u_1}\cdots \bigoplus_{v_{p-1}=u_p}^{u_{p-1}} \bigoplus_{v_p=0}^{u_p}\, [v_1,\ldots,v_p]\,.\label{branchngrylee}\ee
  The result is the set of ground states for the module; among these will be the original field $[s_1,\ldots,s_p]$, which has no derivatives, and the field $[s_1,s_1,\ldots,s_p]$, which has $s_1$ derivatives and is the would-be generalized Weyl tensor for the massless flat-space field.  All the ground states will have $s_1$ columns.  On top of these ground states, we make the excited states by adding more boxes to the top row of each ground state and labelling them with $l\geq s_1$, with $l=s_1$ corresponding to the ground states.

The shift symmetric fields correspond to cutting away the lowest $k+1$ levels, keeping only those states with $l\geq s_1+k+1$.  The field with the lowest number of derivatives among these is the shift invariant field strength \eqref{shiftfieldstrengthe}. 

To describe the PM points, consider first the simplest mixed symmetry field, the massive hook field $[2,1]$.  We add a row of length $2$ to the top to get $[2,2,1]$, and then we branch this down a dimension using \eqref{branchngrylee} to get the states $[2,2,1]$, $[2,2]$, $[2,1,1]$, $[2,1]$.  We can arrange these ground states into a 2 dimensional $2\times 2$ lattice as follows,
\bea  
&& {[2,1]}\ {\rm Weyl\ module\ ground\ states:}\  \raisebox{-45pt}{\epsfig{file=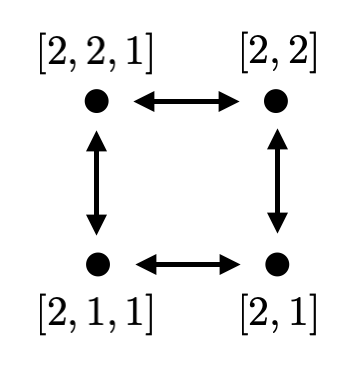,height=1.3in,width=1.3in}}\ \ \ \ ,
 \label{21tanbleagsee}
\eea
with the arrows, which constitute nearest neighbor interactions of the lattice, corresponding to all the possible ways of taking a single derivative and staying within this set.  Adding on the excited states, we then get the full Weyl module of states $[l,2,1]$, $[l,2]$, $[l,1,1]$, $[l,1]$ with $l\geq 2$.  We can imagine these being added onto \eqref{21tanbleagsee} as a third direction in the lattice, coming out of the page, with two-way arrows for all nearest neighbor interactions being added in all directions.  These arrows then give all possible non-vanishing ways of taking derivatives.

There are two PM points for the $[2,1]$ field \cite{Brink:2000ag}, $\underset{\ \ \ \nabla}{[2,1]}$ where the lower square is activated giving a gauge parameter $\underset{\nabla^2}{[2]}$, and $\hspace{-11pt}\underset{\nabla}{\ \ \ [2,1]}$ where the upper square is activated giving a gauge parameter $\underset{0}{[1,1]}$.   
For the $\hspace{-11pt}\underset{\nabla}{\ \ \ [2,1]}$ field, the arrows among the states \eqref{21tanbleagsee} break as follows,
\bea  
&& \underset{\nabla}{\ \ \ [2,1]}\ {\rm Weyl\ module\ ground\ states:}\  \raisebox{-45pt}{\epsfig{file=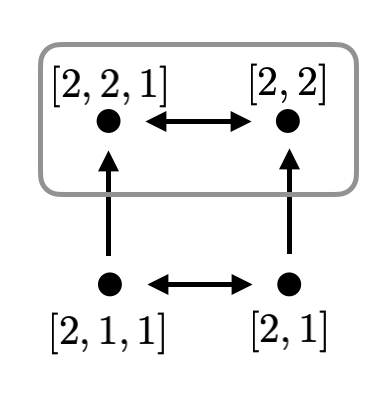,height=1.4in,width=1.4in}} \ \ ,
 \label{21tanbleagsee2}
\eea
with a similar break among all the higher $l$ states.
This leaves a module with $[2,2,1]$ and $[2,2]$ as the new ground states.  The one with the fewest number of derivatives, $[2,2]$, is the PM-invariant field strength \eqref{PMieldstrengthbe}.  
For the $\underset{\ \ \ \nabla}{[2,1]}$ field, the arrows among the states \eqref{21tanbleagsee} break as follows,
\bea  
&& \underset{\ \ \ \nabla}{[2,1]}\ {\rm Weyl\ module\ ground\ states:}\  \raisebox{-40pt}{\epsfig{file=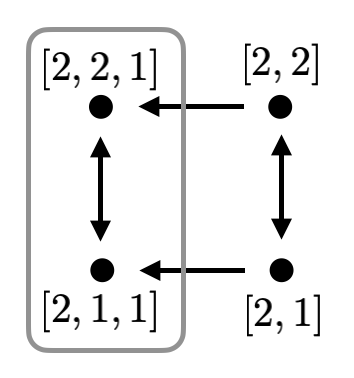,height=1.3in,width=1.3in}}\ \ \ , \nn\\ 
 \label{21tanbleagsee3}
\eea
with a similar break among all the higher $l$ states.
This leaves a module with $[2,2,1]$ and $[2,1,1]$ as the new ground states.  The one with the fewest number of derivatives, $[2,1,1]$, is the PM-invariant field strength \eqref{PMieldstrengthbe}.  

\begin{figure}
\begin{center}
\epsfig{file=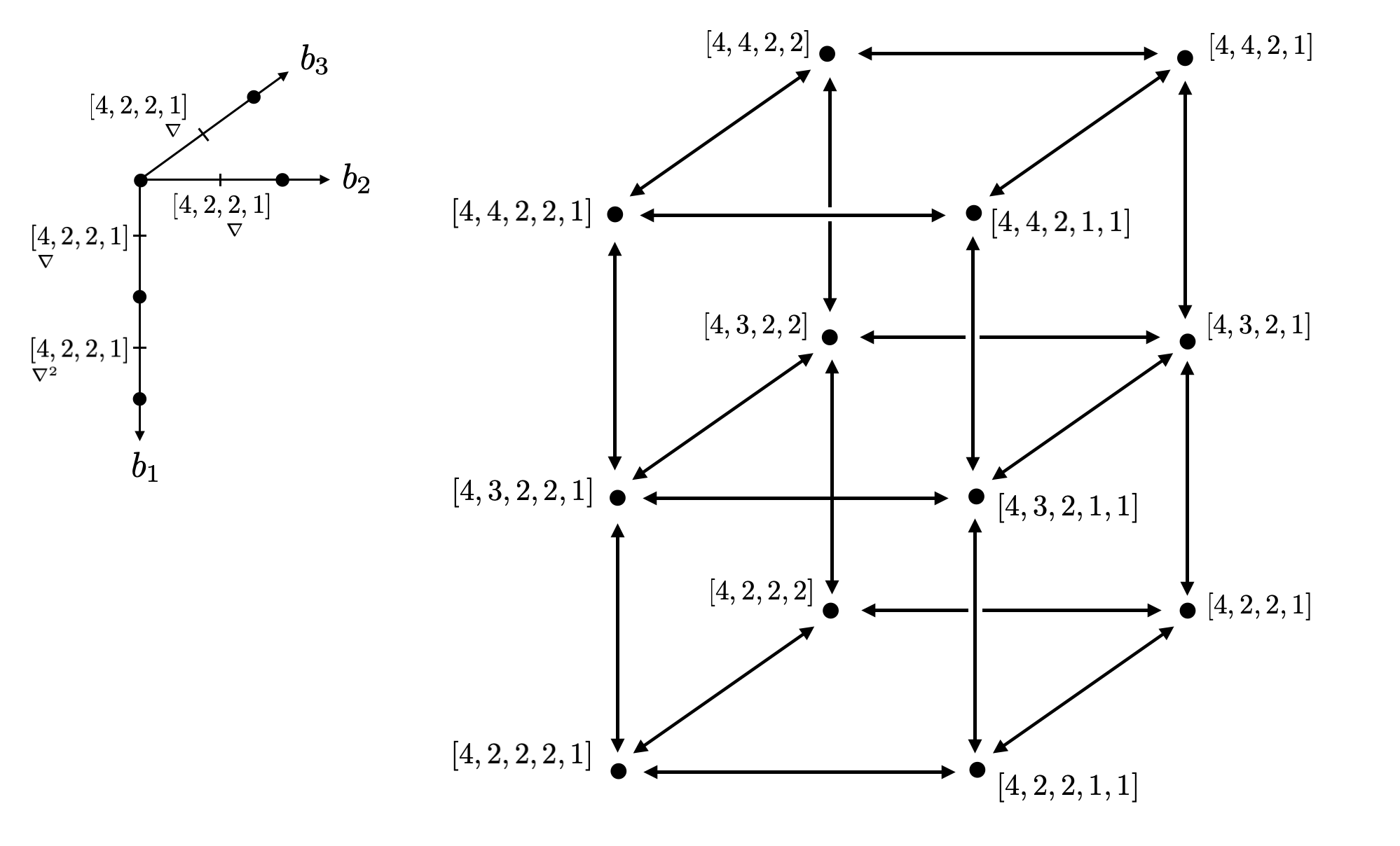,width=6in}
\caption{\small Ground states of the Weyl module for the massive $[4,2,2,1]$ field.
}
\label{4332exmpdee}
\end{center}
\end{figure}

For a more complicated example, consider the $[4,2,2,1]$ field.  Doubling the top row we get $[4,4,2,2,1]$, and then branching down using \eqref{branchngrylee} we get 12 states, which we can arrange into a three dimensional $3\times2\times 2$ lattice as shown in figure \ref{4332exmpdee}.
The highest derivative field $[4,4,2,2,1]$ (the would-be massless field strength) appears in one corner, and the original field $[4,2,2,1]$ appears in the farthest opposite corner.  The remaining states are arranged as follows.  The $[4,2,2,1]$ field has 3 blocks in its tableau: the first block consists of the first row, the second block consists of the next 2 rows, and the third block consists of the final row.  There are three axes, labelled $b_1$, $b_2$, $b_3$ in the figure, corresponding to these 3 blocks.   The number of points along the $I$-th axis ($I=1,2,3$) corresponds to one larger than the number of squares one can remove from the $I$-th block while keeping the tableau legal.  Starting at the corner state $[4,4,2,2,1]$, as we move along the $I$-th axis, we remove at each point a square from the $I$-th block.  

On top of these ground states, we add excited states by adding squares to the first row of each ground state, giving the full set of states $[l,4,2,2,1], \ldots , [l,2,2,1]$ with $l\geq 4$.  We imagine the excited states being added on as a fourth direction in the lattice.  Two way arrows for all the nearest neighbor interactions in the four dimensional lattice give all possible non-vanishing derivatives one can take.

The $[4,2,2,1]$ field has 4 PM points among its 3 blocks: two where the top block is activated, $\underset{\hspace{-33pt} \nabla}{[4,2,2,1]}$, $\hspace{-30pt}\underset{\nabla^2}{\ \ \ \ \ \ \  \ [4,2,2,1]}$, one where the middle block activated, $\underset{\ \ \  \nabla}{[4,2,2,1]}$, and one where the bottom block activated, $\underset{\ \ \ \ \ \ \ \  \nabla}{[4,3,3,1]}$.   Each block axis, $b_I$ labelled in the figure, also shows the PM points occurring between the states, and these PM points correspond to the possible ways of slicing up the ground state lattice through the various PM points: for $\underset{\hspace{-33pt} \nabla}{[4,2,2,1]}$, we cut the downwards arrows leading from the top 4 states, leaving a smaller module consisting of only the top 4 states, the smallest of which is the field strength $[4,4,2,1]$.  For $\underset{\hspace{-33pt} \nabla^2}{[4,2,2,1]}$, we cut the downward arrows leading from the middle 4 states to the bottom 4 states, and we get a smaller module consisting of the top 8 states, the smallest of which is the field strength $[4,3,2,1]$.  For $\underset{\ \ \  \nabla}{[4,2,2,1]}$, we cut the rightward arrows leading from the left 6 states to the right 6 states, giving a smaller module consisting of the left 6 states, the smallest of which is the field strength $[4,2,2,2]$.  For $\underset{\ \ \ \ \ \ \ \  \nabla}{[4,3,3,1]}$, we cut the backwards pointing arrows leading from the front 6 states to the back 6 states, giving a smaller module consisting of the front 6 states, the smallest of which is the field strength $[4,2,2,1,1]$.

In general, for a tableau with $B$ blocks, the lattice of ground states will live in a $B$ dimensional space, and we can label the axes with $b_I$, where $I=1,2,\ldots, B$.  Let $T_I$ be the number of rows in the $I$-th block minus the number of rows in the $(I+1)$-th block, so that $T_I$ is the number of PM points that we can have from activating the $I$-th block.  There will be $T_I+1$ sites in the $b_I$ direction of the lattice, i.e. $b_I=1,2,\ldots ,T_I+1$.   The points in the ground state lattice are filled in as described above: starting with the $[s_1,s_1,\ldots,s_p]$ state in the $b_1=b_2=\cdots=b_B=1$ corner, the rest are filled in by subtracting one square from the $I$-th block for each step in the $I$-th direction, ending up with the original field $[s_1,\ldots, s_p]$ in the far opposite corner $b_1=T_1+1,b_2=T_2+1,\ldots ,b_B=T_B+1$ (this, incidentally, gives an algorithm for computing the branching rule \eqref{branchngrylee}). For the PM point corresponding to activating $t$ squares in the $I$-th block, the states corresponding to the points with $b_I=1,2,\ldots, t$ split off to form the submodule of PM invariant states.  The state with the lowest number of derivatives in this submodule is the PM invariant field strength \eqref{PMieldstrengthbe}.

The missing arrows in a given Weyl module will give the equations of motion satisfied by the operators.  For the PM or shift symmetric fields, the operator with the fewest number of indices, the field strength, will always be completely divergenceless because there is no lower rank tensor for a divergence arrow to go to.   In addition it will satisfy a Bianchi identity given by \eqref{PMieldstrengthbiae} in the PM case and \eqref{shiftfieldstrengthbe} in the shift symmetric case.  This Bianchi identity, divergencelessness, and the Klein-Gordon equation which is satisfied by each state in the module, will then generate all the higher derivative equations of motion implied by the module.  To find the masses in the Klein-Gordon equations, it suffices to recursively use the following identity,
\be \left[ \nabla^2,\nabla_\mu\right] [s_1,\ldots,s_p]=\ldots+ \left(D +1- 2 q  + 2 s_q\right)H^2 [s_1,\ldots,s_q+1,\ldots,s_p]+\ldots \, ,\ee
which tells us that when we act with a derivative and project onto the component of the tableau with one additional square in the $q$-th row, it raises the eigenvalue of $\nabla^2$ by the amount $\left(D +1- 2 q  + 2 s_q\right)H^2$, where $s_q$ is the number of squares in the $q$-th row.

\subsection{Dualities from Weyl modules\label{secweylmoddualsec}}

Having described the Weyl modules, i.e. the sets of on-shell non-trivial local operators linear in the fields and their derivative relations, for all the various types of fields on (A)dS$_D$, we can now use them to understand when dualities exist.  

Each state in the Weyl module is a fully traceless Lorentz tensor in some specific tableau.  These tensors are subject to the rules of permissibility and association reviewed in Appendix \ref{sodrepappendix} for the orthogonal groups.  A tableau is permissible if the sum of the lengths of its first two columns is $\leq D$.  It is not permissible if and only if it has no free components and thus is trivial (like a rank $r$ fully antisymmetric tensor in $D<r$ dimensions).  Two distinct tableaux are {associated} when the length of the first column of the first tableau added to the length of the first column of the second tableau is $D$, and if the remaining columns of the two tableaux are identical.  Two associated tableaux are equivalent as representations of the connected Lorentz group because one can be dualized into the other by contracting the indices of the first column with the $D$ dimensional epsilon tensor.  Among an associated pair, we call the one with fewer indices in its first column the { minimal tensor} and the one with more indices the { non-minimal tensor}.  

A tensor is called {self-associate} if $D$ is even and the length of its first column is $D/2$.  In these cases, the tensor $T$ can be split into self-dual and anti-self-dual parts $T^\pm$ with respect to its first-column indices.  We call $T^\pm$ the chiral halves of $T$.  Due to the Lorentzian relation $\ast^2=(-1)^{D/2+1}$ for the Hodge star acting on $(D/2)$-forms, the self-duality is real for $D/2$ odd,
\be T^\pm={1\over 2}\left(T\pm \ast T\right),\ \ \ \ast T^\pm=\pm T^\pm,\ \ \ T=T^++T^-,\ee
(where the indices on the tensor $T$ are suppressed and the Hodge star operator acts only on the indices of the first column)
 and imaginary for $D/2$ even,
\be T^\pm={1\over 2}\left( T\mp i\ast T\right)\,, \ \ \ast T^\pm=\pm i \,T^\pm\,,\ \ \ T=T^++T^-\,. \ee

A duality between two fields exists if their Weyl modules are equivalent upon removing non-permissible tableaux (which vanish identically and thus are not part of the Weyl module), dualizing all non-minimal tensors to their minimal counterparts, and splitting up all self-associated tableau into chiral halves.  A self-duality exists if a field's Weyl module splits into two irreducible parts upon the same simplification.  In this subsection, we go through some examples of how this happens, before describing the general pattern of dualities in the following sections.

\textbf{Massless $p$-forms:}  Consider first the best known examples of duality, those between two massless $p$-forms each with $p\geq 1$.  A massive $p$-form $a=[1^p]$, $1\leq p\leq D-3$, with mass $\tilde m_{[1^{p}]}$, has a two-row Weyl module, where the ground states are the $p$-form itself, and its $(p+1)$-form field strength $[1^{p+1}]\propto da$, 
\bea && [1^p]\ {\rm Weyl\ module:}\ \ \  \raisebox{-46pt}{\epsfig{file=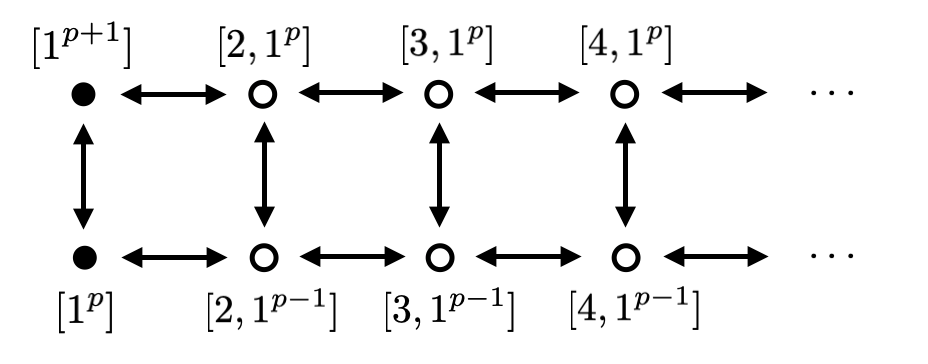,height=1.3in,width=3.2in}}\ \ \ .
\label{pformweylme}
\eea
The ground states satisfy the Klein-Gordon equations
\bea
&&\left(\nabla^2 -\tilde m_{[1^{p}]}^2-(D-2p-1)H^2\right) [1^{p+1}]=0\, , \label{massivepformewe1} \\
&&\left(\nabla^2 -\tilde m_{[1^p]}^2\right) [1^{p}]=0 \,.\label{massivepformewe2}
\eea

When the mass takes the value $\tilde m_{\underset{\nabla}{[1^p]}}^2=p(D-p)H^2$, the $p$-form is massless (its only PM point), and the first row of \eqref{pformweylme} separates off into an irreducible module, with its ground state $[1^{p+1}]$ proportional to the gauge invariant field strength $da$, satisfying
\be \left(\nabla^2 -(p+1)(D-p-1)H^2\right) [1^{p+1}]=0\,.\label{pp1feme}\ee

If we dualize this one-row Weyl module by contracting the first column of each state with the $D$ dimensional epsilon tensor, we get a module built from the ground state $[1^{D-p-1}]$, satisfying a Klein-Gordon equation with the same mass as \eqref{pp1feme}.  This is precisely the module for a massless $D-p-2$ form $b$, with ground state proportional to the $(D-p-1)$-form field strength $db$.  The basic duality relation relates these two ground states,
\be db=\ast da\, ,\ \ \ da=-(-1)^{(p+1)(D-p-1)}\ast db\, .\label{masslesspdobdree}\ee
The normalization of the first of these relations is arbitrary, the normalization of the second is fixed by requiring the two relations to be consistent  with each other (using the Lorentzian identity $\ast^2=-(-1)^{p(D-p)}$ when acting on a $p$ form).

In the case where $D$ is even and $p={D\over 2}-1$, the massless $p$-form can be dual to itself.  We then then break up the massless module by splitting each state into self-dual and anti-self-dual parts with respect to the indices in the first row of its tableau:  
\begin{itemize}
\item  For $D=2,6,10,\ldots$, we have $\ast^2=1$ when acting on $(D/2)$-forms, and we have real self-dual and anti-self-dual parts.  The splitting for the field strength in the ground state reads
\be da^\pm={1\over 2}\left(da\pm \ast da\right)\,,\ \ \ \ast da^\pm=\pm da^\pm\,,\ \ \ da=da^++da^-\,. \label{masslepformfre}\ee
\item  For $D=4,8,12,\ldots$, we have $\ast^2=-1$ when acting on $(D/2)$-forms, and we have imaginary self-dual and anti-self-dual parts. The splitting for the field strength in the ground state reads
\be da^\pm={1\over 2}\left( da\mp i\ast da\right)\,, \ \ \ast da^\pm=\pm i \,da^\pm\,,\ \ \ da=da^++da^-\,. \label{masslepformfre2}\ee
\end{itemize}
By acting with any of the derivatives expressed via arrows in the massless module, we cannot go from a self-dual to an anti-self-dual state or vice versa, so the module splits into two separate irreducible modules, one corresponding to each choice of sign in \eqref{masslepformfre} or \eqref{masslepformfre2}, which encode the chiral halves of the massless $(D/2-1)$-form.  We call these chiral halves $\underset{\hspace{-6pt}\nabla}{[1^p]_\pm}$.

\textbf{$(D-2)$-form and $k=0$ scalar:}
For $p=D-2$, the massless $p$-form field has the single-row Weyl module with ground state the field strength $[1^{D-1}]$, satisfying 
\be \left(\nabla^2 -(D-1)H^2\right) [1^{D-1}]=0\,.\label{pp1fem2e}\ee

This states in this module dualize into one built on a 1-form $[1]$ satisfying the same equation.  This is the same module as the $k=0$ shift symmetric scalar; the masses \eqref{sckpp1fem2e} and \eqref{pp1fem2e} match for $k=0$.  Note that this takes into account the fact that the basic scalar $\phi$ is not included in the duality; it must be factored out via the $k=0$ shift symmetry.  It is in this sense that we can think of the massless scalar as a massless 0-form.  The basic duality relation relates the gauge invariant field strength of the $p$-form field with the shift invariant field strength of the scalar,
\be d\phi=\ast da\, ,\ \ \ da=(-1)^{D}\ast d\phi\, .\ee

In the degenerate case $D=2$, the massless scalar is dual to itself, and its module built from the field strength $d\phi$ can be split into real self-dual and anti-self-dual parts, which are the chiral and anti-chiral boson (a similar split for the higher $k$ shift symmetric scalars in $D=2$ is discussed below),
\be d\phi^\pm={1\over 2}\left( d\phi\pm \ast d\phi\right)\,,\ \ \ \ast d\phi^\pm=\pm d\phi^\pm\,,\ \ \ d\phi=d\phi^++d\phi^-\,. \ee

\textbf{Massive $p$-forms:}  Consider a massive $p$-form $a$, $1\leq p\leq D-2$, with mass $\tilde m_{[1^{p}]}$.  It has a two-row module shown in \eqref{pformweylme} generated by $a=[1^p]$, $da\propto [1^{p+1}]$ satisfying \eqref{massivepformewe1}, \eqref{massivepformewe2}.  If we dualize both rows, we get a module generated by $[1^{D-p}]$ and $[1^{D-p-1}]$.  This is the Weyl module content of a massive $(D-p-1)$-form, $b$, with the masses related as
\be \tilde m_{[1^{D-p-1}]}^2= \tilde m_{[1^{p}]}^2+(D-2p-1)H^2.\label{massipfmassee}\ee 
These masses are such that the dual CFT conformal dimensions in \eqref{mixsymmassde2ree} of the two fields are the same.

The basic duality relation among the fields is read off from the identifications of the ground states of the modules on each side of the duality,
\bea && [1^p]_{a}\propto \ast [1^{D-p}]_{b} \quad\quad \ \hspace{6pt} \Rightarrow\ \   a=\ast db\, , \label{cmgreled1}\\
   && [1^{D-p-1}]_{b} \propto\ast [1^{p+1}]_{a} \quad  \Rightarrow \ \  b={(-1)^{Dp+1}\over \tilde m_{[1^p]}^2 -p(D-p)H^2}\ast da\,. \label{cmgreled2}
\eea
The normalization in \eqref{cmgreled1} is set arbitrarily, and the normalization in \eqref{cmgreled2} is fixed by requiring that \eqref{cmgreled1} and \eqref{cmgreled2} are consistent with each other (which requires using the on shell conditions $d^\dag a=d^\dag b=0$ as well as the Klein-Gordon equations).  Note that the value $ \tilde m_{[1^p]}^2 =p(D-p)H^2$ for which this normalization is singular is the massless value for the $b$ field.

When $D$ is odd and $2p+1=D$, the two rows of the massive module \eqref{pformweylme} can be made to have identical shapes by dualizing the top row, and the masses in \eqref{massipfmassee} are equal.  This massive $p$-form in this case is dual to itself, with the relation $a\propto \ast da$.
The module will therefore split into chiral halves.  To affect this split, first note that the operator $\ast d$ maps the two ground states of the Weyl module into each other once the second row is dualized, so it maps the space of on-shell $p$-forms into itself, preserving the Klein-Gordon equation and on-shell condition $d^\dag a=0$.  It satisfies $(\ast d)^2=(-1)^{p+1}\left(   \tilde m_{[1^p]}^2 -p(p+1)H^2\right)$ on-shell, thus we can define an operator ${\cal P}\equiv {1\over \sqrt{ \tilde m_{[1^p]}^2 -p(p+1)H^2}}\ast d$ which satisfies ${\cal P}^2=1$ for $p$ odd and ${\cal P}^2=-1$ for $p$ even, and use it to project as follows:
\begin{itemize}
\item  For $p$ odd ($D=3,7,11,\ldots$), we take 
\be a^\pm={1\over 2}\left(a\pm {\cal P} a\right),\ \ \ {\cal P} a^\pm=\pm a^\pm,\ \ \ a=a^++a^-\, ,  \label{massivpfsse1} \ee 
and the duality relation is real.
\item  For $p$ even ($D=5,9,13,\ldots$), we take 
\be a^\pm={1\over 2}\left(a\mp i {\cal P} a\right),\ \ \ {\cal P} a^\pm=\pm i a^\pm,\ \ \ a=a^++a^-\, ,  \label{massivpfsse2} \ee 
and the duality relation is imaginary.
\end{itemize}
(Note that we have implicitly assumed the field is unitary so that the mass squared is above its massless value, $\tilde m_{[1^p]}^2 >p(p+1)H^2$.  If the mass squared  is below its massless value, then the dimensions in which the duality is real and imaginary are reversed.) 
The module splits into two separate irreducible modules, one corresponding to each choice of sign in \eqref{massivpfsse1} or \eqref{massivpfsse2}.  These encode the chiral halves of the massive $\left({D-1\over 2}\right)$-form,  which we call $[1^p]_\pm$.
Massive $p$-form dualities and self-dualities have been studied in \cite{Townsend:1981nu,Townsend:1983xs,Cecotti:1987qr,Casini:2002jm,Zinoviev:2005zj,Buchbinder:2008jf,Buchbinder:2009pa,Dalmazi:2011df,Kuzenko:2020zad,Barbosa:2022zfm}.  

These massive dualities also relate the shift symmetric $p$-forms of level $k$ to each other.  The modules for the level $k$ shift fields are simply missing the first $k+1$ columns, but the rest of the columns dualize in the same way and the masses of the $k$-th level are the same on both sides of the duality.  The basic duality relation now relates the shift symmetric field strength \eqref{shiftfieldstrengthe} on one side to an exterior derivative acting on the first column of the field strength of the other: the field strength for $\underset{k}{[1^p]}$ has the shape $[k+2,1^{p-1}]$, the field strength for $\underset{k}{[1^{D-p-1}]}$ has the shape $[k+2,1^{D-p-2}]$, and these two fields strengths are dual to each other in a manner analogous to \eqref{cmgreled1}, \eqref{cmgreled2}, or  \eqref{massivpfsse1}, \eqref{massivpfsse2} in the self-dual case.

\textbf{Massive $D-1$ form and scalar:}
A massive $D-1$ form $a$ has a module generated by the ground states $a= [1^{D-1}]$ and $da\propto [1^{D}]$.  Dualizing the states generated by $[1^{D-1}]$ creates a set of states generated by $[1]$.  Of all the states generated by $[1^{D}]$, all of them are trivial except for the ground state itself, which is dual to a scalar $[0]$.  Adding this scalar to the states generated by $[1]$ gives a set generated by $[0]$, equivalent to the Weyl module of a massive scalar $\phi$, with the masses related as
\be \tilde m_{[1^{D-1}]}^2= \tilde m_{[0]}^2+(D-1)H^2.\label{massipfmassesce}\ee 
The basic duality relation is
\bea   \phi=\ast da\, ,\ \ \   a=-{1\over \tilde m_{[0]}^2}\ast d\phi\,. \label{masslesspdobdreee2e}
\eea
Dualities of this type are studied in \cite{Curtright:1980yj,Curtright:2019yur,Barbosa:2022zfm}

This scalar/massive $(D-1)$-form duality also holds for the shift-symmetric fields.  In this case the $(D-1)$-form at level $k$ is dual to a scalar at level $k+1$, as can be seen from the fact that the modules are identical and the masses the same on both sides.  The basic duality relation relates the shift symmetric field strength \eqref{shiftfieldstrengthe} on one side to the field strength of the other: the field strength for $\underset{k}{[1^{D-1}]}$ has the shape $[k+2,1^{D-2}]$, the field strength for $\underset{k+1}{[0]}$ has the shape $[k+2]$, and these two fields strengths are dual to each other in a manner analogous to \eqref{masslesspdobdreee2e}.

This relation degenerates in $D=2$, and we have self-duality among the shift symmetric scalars and vectors. The states  of the level $k$ shift symmetric scalar in $D=2$ are the symmetric tensors built on the ground state $[k+1]=\nabla_{(\mu_1}\cdots \nabla_{\mu_{k+1})_T}\phi\equiv C_{\mu_1\ldots \mu_{k+1}}$.  In $D=2$, symmetric traceless tensors are dual to themselves when we dualize along the first index, and we can form the real self-dual and anti-self-dual combinations as follows
\bea && C_{\mu_1\ldots \mu_{k+1}}^\pm={1\over 2}\left(C_{\mu_1\cdots \mu_{k+1}}\pm \epsilon^\nu_{\ \mu_1} C_{\nu\mu_2\cdots \mu_{k+1}} \right)\,,  \nn \\
&& \epsilon^\nu_{\ \mu_1} C_{\nu\mu_2\ldots \mu_{k+1}}^\pm=\pm C_{\mu_1\ldots \mu_{k+1}}^\pm\, \ \ \  C_{\mu_1\ldots \mu_{k+1}}=C_{\mu_1\ldots \mu_{k+1}}^++C_{\mu_1\ldots \mu_{k+1}}^- \,.  \label{shiftscalesee2}
\eea
Thus the shift symmetric scalar breaks up into chiral halves; this is well-known for the massless scalar, which is the case $k=0$.  The shift symmetric scalars in $D=2$ are also known as discrete series fields and have been studied in \cite{Joung:2007je,Anninos:2023lin}.
Note that the generic massive scalar does not split into chiral parts like this: the presence of the scalar itself as the $[0]$ state spoils the duality because this state cannot be split into chiral halves.

The same self-duality occurs for the shift-symmetric vectors in $D=2$.  The level $k$ shift symmetric vector is dual to the level $k+1$ shift symmetric scalar, and thus has the same module content which can be split in the same way as in \eqref{shiftscalesee2}.

\textbf{Massive spin $s$ self-duality in $D=3$:}
Consider a spin $s\geq 1$ field $h_{\mu_1\cdots\mu_s}$ with mass $\tilde m_{[s]}$.  Its Weyl module is pictured in \eqref{massivevmoduleespins2}, and the ground states in the lowest two rows satisfy the Klein-Gordon equations
\bea && \left(\nabla^2 -\tilde m_{[s]}-(D-3)H^2\right) [s,1]=0\, , \nn\\
&&  \left(\nabla^2 -\tilde m_{[s]}^2 \right) [s]=0\, . \label{massiveeD3kgee}
\eea 
 
 In $D=3$, all the rows in the Weyl module except for the bottom two trivialize, and the states in the second become equal in shape to those in the first row when we dualize along their first column indices. The masses in the Klein-Gordon equations \eqref{massiveeD3kgee} for the two rows also become equal only in $D=3$.  The field has a self-duality in $D=3$ that relates the two rows $[s]\propto \ast [s,1]$, and the module splits up into chiral halves that we can identify as follows.  Define the operator 
 \be {\cal P}h_{\mu_1\ldots \mu_s}\equiv {1\over \sqrt{\tilde m_{[s]}^2-(s+1)H^2}} \epsilon_{\mu_1\nu_1\nu_2}\nabla^{\nu_1}h^{\nu_2}_{\ \ \mu_2\cdots \mu_s}\,.\ee
It maps the two ground states of the Weyl module into each other once the second row is dualized, so it maps the space of on-shell symmetric tensors into itself, preserving the Klein-Gordon equation and the conditions of divergencelessness and tracelessness. 
The mass $\tilde m_{[s]}^2=(s+1)H^2$ where the denominator vanishes is the maximal depth ($t=s$) PM value, which also sets the Higuchi bound \cite{Higuchi:1985ad} for unitarity.  As long as the mass is above this, the operator satisfies ${\cal P}^2=1$ and we can use it to project onto real self-dual and anti-self-dual parts as follows:
\be h^\pm={1\over 2}\left(h\pm {\cal P} h\right),\ \ \ {\cal P} h^\pm=\pm h^\pm,\ \ \ h=h^++h^-\, ,  \label{massivpfshse1} \ee 
where we have suppressed the indices on $h$.
(Note that if the mass squared is below the Higuchi bound, we would have ${\cal P}^2=-1$ and would have imaginary self-duality relations analogous to \eqref{massivpfsse2}.)  The module splits into two separate irreducible modules, one corresponding to each choice of sign in \eqref{massivpfshse1}. These encode the chiral halves of the massive spin $s$ field,  which we call $[s]_\pm$.   Topologically massive gravity, when expanded around an (A)dS background, is an example of a theory that propagates only one of the two chiral halves for $s=2$ \cite{Carlip:2008eq,Carlip:2008jk}.

\textbf{Maximal depth PM and shift symmetric scalars in $D=3$:}
Consider a spin $s$ maximal depth ($t=s$) PM field.  Of the $s+1$ rows in the Weyl module \eqref{massivevmoduleespins2} for a generic mass spin $s$ field, generated by the ground states $[s,s],\ [s,s-1],\ldots,[s,1],\ [s]$, the last one, generated by $[s]$, is removed in the case of the maximal depth PM field.  
The $[s,1]$ state is the field strength \eqref{shiftfieldstrengthe} and it satisfies the Klein-Gordon equation
\be \left(\nabla^2 -(s+2D-5)H^2\right) [s,1]=0\,.\label{PMD3kgee}\ee

In $D=3$, the states generated by $[s,s],\ [s,s-1],\ldots,[s,2]$ are all trivial, and those generated by $[s,1]$ can be dualized into a set generated by $[s]$.  This is precisely the set of fields for the $k=s-1$ shift symmetric scalar in $D=3$, and the masses in \eqref{PMD3kgee} agree with those in \eqref{sckpp1fem2e} precisely in $D=3$: the maximal depth spin $s$ PM field is dual to the level $k=s-1$ shift symmetric scalar.

The duality relation is given through the duality between the field strength $[s,1]= 2\nabla_{[\nu_1}h_{\nu_2]\mu_2\cdots\mu_{s-1}}\equiv F_{\nu_1\nu_2\mu_1\ldots \mu_{s-1}}$ of the PM field $h_{\mu_1\cdots \mu_s}$, and the field strength $[s]=\nabla_{(\mu_1}\cdots\nabla_{\mu_s)_T}\phi\equiv C_{\mu_1\ldots \mu_s}$ of the $k=s-1$ scalar $\phi$,
\be C_{\mu_1\ldots \mu_s}={1\over 2}\epsilon_{\nu_1\nu_2\mu_1}F^{\nu_1\nu_2}_{\ \ \ \ \ \mu_2\cdots\mu_s}, \ \ \  F_{\nu_1\nu_2\mu_1\ldots \mu_{s-1}}=-\epsilon_{\nu_1\nu_2\alpha}C^\alpha_{\ \ \mu_1\cdots\mu_{s-1}} \,.\ee

The case $s=1$ is the well-known statement that the photon is dual to a massless scalar in $D=3$ (also covered by the scalar/$p$-form discussion above).  The case $s=2$ says that the PM graviton is dual to a $k=1$ scalar in $D=3$, which was studied in \cite{Galviz:2017tda} (the $k=1$ scalar is also known as the (A)dS galileon \cite{Goon:2011qf,Goon:2011uw,Burrage:2011bt,Bonifacio:2021mrf}).

\textbf{Curtright field and massive graviton in $D=4$:}  Consider a massive $[2,1]$ field $c_{\mu\nu,\rho} \in \raisebox{5pt}{\resizebox{.5cm}{!}{\gyoung(\mu\rho ,\nu )}}$ with a mass $\tilde m_{[2,1]}$, known as a Curtright field \cite{Curtright:1980yk,Curtright:1980yj}.  This has a Weyl module generated by the ground states $[2,2,1]$, $[2,1,1]$, $[2,2]$, $[2,1]$, as shown in \eqref{21tanbleagsee}.  These ground states satisfy the Klein-Gordon equations
\bea
&&\left(\nabla^2 -\tilde m_{[2,1]}^2-2(D-3)H^2\right) [2,2,1]=0\, ,\\ 
&&\left(\nabla^2 -\tilde m_{[2,1]}^2-(D-5)H^2\right) [2,1,1]=0\, ,\\
&&\left(\nabla^2 -\tilde m_{[2,1]}^2-(D-1)H^2\right) [2,2]=0\, ,\\
&&\left(\nabla^2 -\tilde m_{[2,1]}^2\right) [2,1]=0\, .
\eea

Now consider a massive graviton, i.e. a $[2]$ field $h_{\mu\nu}$, with a mass $\tilde m_{[2]}$.  This has a Weyl module with the ground states $[2,2]$, $[2,1]$, $[2]$ which satisfy the Klein-Gordon equations
\bea
&&\left(\nabla^2 -\tilde m_{[2]}^2-2(D-2)H^2\right) [2,2]=0\, ,\\ 
&&\left(\nabla^2 -\tilde m_{[2]}^2-(D-3)H^2\right) [2,1]=0\, ,\\
&&\left(\nabla^2 -\tilde m_{[2]}^2\right) [2]=0\, .
\eea

In $D=4$, of the states in the Curtright field module, those built on $[2,2,1]$ vanish identically and those built on $[2,1,1]$ can be dualized to states built on a $[2]$.  Thus in $D=4$, the module for the Curtright field and the module for the massive graviton contain the same field types.  Furthermore, the masses in all the Klein-Gordon equations agree, only in $D=4$, if we make the identification
\be \tilde m_{[2,1]}^2= \tilde m_{[2]}^2+H^2\,. \label{pmcurtmarele}\ee
According to the AdS/CFT mass formula \eqref{mixsymmassde2ree}, this mass relation means that both fields carry the same dual conformal dimension.

The basic duality relation among the fields is read off from those identifications of the ground states that involve the bare fields,
\bea  &&[2,1]_{\rm Curtright}=\ast [2,1]_{\rm graviton} \Rightarrow  c_{\mu\nu,\rho}={1\over 2} \epsilon_{\mu\nu\alpha\beta}\nabla^\alpha h^{\beta}_{\ \rho}, \label{cmgrele1}\\
&& [2]_{\rm graviton}\propto\ast [2,1,1]_{\rm Curtright}  \Rightarrow  h_{\mu\nu}={1\over \tilde m_{[2,1]}^2-5H^2} \epsilon_{\mu\alpha\beta\gamma} \nabla^\alpha c^{\beta\gamma}_{\ \ \ \nu} \,. \label{cmgrele2}
\eea
The normalization in \eqref{cmgrele2} is fixed by requiring that \eqref{cmgrele1} and \eqref{cmgrele2} are consistent with each other on shell.  (Note that the value $\tilde m_{[2,1]}^2=5H^2$ for which this normalization is singular is the $\underset{\ \ \  \nabla}{[2,1]}$ PM value for the $D=4$ Curtright field.)  
The remaining relation $[2,2]_{\rm graviton}=\ast [2,2]_{\rm Curtright}$, involving derivatives on both sides, follows by taking a further derivative of \eqref{cmgrele1} and Young symmetrizing over $[2,2]$.

In higher $D$, it can be seen from similar manipulations that the massive graviton is dual to a massive $[2,D-3]$ field \cite{Gonzalez:2008ar,Khoudeir:2008bu,Alshal:2019hpk}.

\textbf{PM hook and massless graviton in $D=4$:}
Consider the mass $\tilde m_{[2,1]}^2=3H^2$ in arbitrary $D$, for which the $[2,1]$ field gets the PM symmetry $\underset{\hspace{-12pt}\nabla}{[2,1]}$ where the top block is activated.   This has a Weyl module generated by the ground states $[2,2,1]$, $[2,2]$, as shown in \eqref{21tanbleagsee2}.  These ground states satisfy the Klein-Gordon equations
\bea
&&\left(\nabla^2 -(2D-3)H^2\right) [2,2,1]=0\, ,\\ 
&&\left(\nabla^2 -(D+2)H^2\right) [2,2]=0\, . \label{22hookpmkged1e}
\eea

Now consider a massless graviton in $D$ dimensions, i.e. a $\underset{\nabla}{[2]}$ field $h_{\mu\nu}$ with a mass value $\tilde m_{\underset{\nabla}{[2]}}^2=2H^2$.  This has a Weyl module with the ground states $[2,2]$, which satisfies the Klein-Gordon equation
\bea
&&\left(\nabla^2 -2(D-1)H^2\right) [2,2]=0\, . \label{2mpmkhe2ee}
\eea

In $D=4$, the mass relation \eqref{pmcurtmarele} equates the massless graviton mass with the $\underset{\hspace{-12pt}\nabla}{[2,1]}$ mass.  The states in the Curtright field module built on $[2,2,1]$ vanish identically, leaving only the $[2,2]$ states, which match those of the massless graviton, and the masses in the Klein-Gordon equations \eqref{22hookpmkged1e}, \eqref{2mpmkhe2ee} also agree only in $D=4$.
The basic duality relation now links the two PM field strengths with each other,
\bea  &&[2,2]_{\rm Curtright}=\ast [2,2]_{\rm graviton} \Rightarrow  F_{\mu\nu\rho\sigma}={1\over 2} \epsilon_{\mu\nu\alpha\beta}C^{\alpha\beta}_{\ \ \ \rho\sigma}\Rightarrow C_{\mu\nu\rho\sigma}=-{1\over 2} \epsilon_{\mu\nu\alpha\beta}F^{\alpha\beta}_{\ \ \ \rho\sigma} , \nn\\ \label{cmongrele1}
\eea
where $C_{\mu\nu\rho\sigma}\equiv {\cal Y}^T_{\resizebox{.5cm}{!}{\gyoung(\mu\rho ,\nu\sigma )}}\nabla_\mu\nabla_\rho h_{\nu\sigma}$ is the field strength for the massless graviton (proportional to the linearized Weyl tensor) and $F_{\mu\nu\rho\sigma}\equiv {\cal Y}^T_{\resizebox{.5cm}{!}{\gyoung(\mu\rho ,\nu\sigma )}}\nabla_\sigma C_{\mu \nu\rho}$ is the PM field strength for the $\underset{\hspace{-12pt}\nabla}{[2,1]}$ PM field.

Since the massless graviton possesses a (imaginary) self-duality in $D=4$ where the Weyl tensor is transformed into its dual,  the $\underset{\hspace{-12pt}\nabla}{[2,1]}$ field also has a similar self-duality, where its field strength is transformed into its dual.

In $D>4$, similar manipulations show that the massless graviton is dual to a $\underset{\hspace{-30pt}\nabla}{[2,1^{D-3}]}$ partially massless hook \cite{Basile:2015jjd}, though in these cases there is no self-duality on either side.

\textbf{PM hook and galileon scalar in $D=4$:}
Now consider the mass $\tilde m_{[2,1]}^2=(2D-3)H^2$, for which the $[2,1]$ field on (A)dS gets the PM symmetry $\underset{\ \ \ \nabla}{[2,1]}$ with the bottom block activated \cite{Brink:2000ag}.   This has a Weyl module generated by the ground states $[2,2,1]$, $[2,1,1]$, as shown in \eqref{21tanbleagsee3}.  These ground states satisfy the Klein-Gordon equations
\bea
&&\left(\nabla^2 -(4D-9)H^2\right) [2,2,1]=0\, ,\\ 
&&\left(\nabla^2 -(3D-8)H^2\right) [2,1,1]=0\, . \label{2211mksmae}
\eea

Now consider the scalar in $D$ dimensions with mass taking the $k=1$ shift symmetric value $\tilde m_{\underset{1}{[0]}}^2=-DH^2$.  The lowest two states, $[0]$, $[1]$, of the scalar module are removed by the shift symmetry, so the $k=1$ scalar $\underset{1}{[0]}$ has a Weyl module with the ground state $[2]$, which satisfies the Klein-Gordon equation \eqref{sckpp1fem2e} with $k=1$,
\be \left(\nabla^2 -DH^2\right) [2]=0.\label{scak1mae}\ee

In $D=4$, the states built on $[2,2,1]$ in the $\underset{\ \ \ \nabla}{[2,1]}$ module vanish identically and those built on $[2,1,1]$ can be dualized to a set built on $[2]$.  Furthermore, the mass in the Klein-Gordon equation \eqref{scak1mae} agrees with that in \eqref{2211mksmae} when $D=4$.   The $\underset{\ \ \ \nabla}{[2,1]}$ field and the $\underset{1}{[0]}$ field have the same Weyl module in $D=4$ and are dual to each other.

The basic duality relation links the PM field strength of the $\underset{\ \ \ \nabla}{[2,1]}$  field with the shift invariant field strength of the $k=1$ scalar, 
\bea  &&[2,1,1]_{\rm Curtright}=\ast [2]_{\rm scalar} \Rightarrow  F_{\mu\nu\rho\sigma}= \epsilon_{\beta\mu\nu\rho}C^{\beta}_{\ \ \sigma}\Rightarrow C_{\mu\nu}=-{1\over 3!} \epsilon_{\mu\alpha\beta\gamma}F^{\alpha\beta\gamma}_{\ \ \ \ \nu} , \nn\\ \label{cmongrele1}
\eea
where $C_{\mu\nu}=\nabla_{(\mu}\nabla_{\nu)_T}\phi$ is the field strength for the $k=1$ scalar and $F_{\mu\nu\rho\sigma}=\nabla_{[\mu} c_{\nu\rho]\sigma}$ is the PM field strength for the $\underset{\ \  \ \nabla}{[2,1]}$ PM point.

Through similar arguments, we can see that in general $D$, the level $k$ shift symmetric scalar $\underset{k}{[0]}$ is dual to the long hook PM field $\underset{\ \ \ \ \ \ \ \  \nabla}{ [k+1,1^{D-3}]}$ where the bottom block is activated; the $k=0$ case is the duality between the massless scalar and the massless $(D-2)$-form, already discussed above.

\textbf{Graviton triality in $D=5$:}  
Consider the PM hook field $\underset{\hspace{-22pt}\nabla}{[2,1,1]}$ in which the square in the top row is activated, which occurs at the mass value $\tilde m_{\underset{\hspace{-14pt}\nabla}{[2,1,1]}}^2=4H^2$ in any $D$.  The ground states of the Weyl module are $[2,2,1,1]$ and $[2,2,1]$, which satisfy the Klein-Gordon equations
\bea
&&\left(\nabla^2 -(2D-4)H^2\right) [2,2,1,1]=0\, , \label{gravtir211eqetope}  \\ 
&&\left(\nabla^2 -(D+3)H^2\right) [2,2,1]=0\, . \label{gravtir211eqe}
\eea
The state $[2,2,1]$ is the field strength \eqref{PMieldstrengthbe}.

Now consider the PM window field $\underset{\hspace{14pt}\nabla^2}{[2,2]}$ in which both of the bottom squares are activated, which occurs at the mass value $\tilde m_{\underset{\hspace{10pt}\nabla^2}{[2,2]}}^2=2(D-1)H^2$.  The ground states of the Weyl module are $[2,2,2]$ and $[2,2,1]$, which satisfy the Klein-Gordon equations
\bea
&&\left(\nabla^2 -2(2D-5)H^2\right) [2,2,2]=0\, , \label{gravtir211eqetope2} \\ 
&&\left(\nabla^2 -(3D-7)H^2\right) [2,2,1]=0\, . \label{gravtir211eqe2}
\eea
The state $[2,2,1]$ is the field strength \eqref{PMieldstrengthbe}.

Note that these two fields have a field strength with the same shape $[2,2,1]$, but in general $D$ they are different fields because they have different masses in their Klein-Gordon equations \eqref{gravtir211eqe}, \eqref{gravtir211eqe2}, and the top row of the Weyl modules, \eqref{gravtir211eqetope}, \eqref{gravtir211eqetope2}, are different.  (They also satisfy different Bianchi identities \eqref{PMieldstrengthbiae}: $[2,2,2]$ for the $\underset{\hspace{-22pt}\nabla}{[2,1,1]}$ field and $[2,2,1,1]$ for the $\underset{\hspace{14pt}\nabla^2}{[2,2]}$ field.)

However in $D=5$, the top rows \eqref{gravtir211eqetope}, \eqref{gravtir211eqetope2} of the Weyl modules trivialize, and the masses in the Klein-Gordon equations \eqref{gravtir211eqe}, \eqref{gravtir211eqe2} become equal.  These two fields are now dual to each other, both with a single-row Weyl module whose ground state $[2,2,1]$  satisfies
\bea
&&\left(\nabla^2 -8H^2\right) [2,2,1]=0\, . \label{gravtir211eqe3}
\eea
Note that this duality does not involve dualizing one of the field strengths with an epsilon symbol; the two field strengths are directly equated with each other. (Note also that the Bianchi identities trivialize in $D=5$: both fields satisfy both Bianchi identities.)

If we dualize the states generated by $[2,2,1]$, we get a set generated by $[2,2]$ with the same Klein-Gordon mass as in \eqref{gravtir211eqe3}.  This is now the same as the Weyl module \eqref{2mpmkhe2ee} for a massless graviton $\underset{\nabla}{[2]}$; the Klein-Gordon masses agree precisely when $D=5$.  We have a triality: both the $\underset{\hspace{-22pt}\nabla}{[2,1,1]}$ field and the $\underset{\hspace{14pt}\nabla^2}{[2,2]}$ field are dual to the massless graviton, with a duality that relates the field strengths through the epsilon symbol, whereas $\underset{\hspace{-22pt}\nabla}{[2,1,1]}$  and  $\underset{\hspace{14pt}\nabla^2}{[2,2]}$ are dual to each other with a duality that relates the field strengths directly without an epsilon symbol.

In $D>5$ there is also a similar triality with two dual gravitons: the dual gravitons are $\underset{\hspace{-30pt}\nabla}{[2,1^{D-3}]}$ and $\underset{\hspace{23pt}\nabla}{[2,2,1^{D-5}]}$.  Note that this phenomenon of having two dual gravitons does not occur in flat space.  As reviewed in Appendix \eqref{masslessdflatsection}, in flat space the dual to the graviton occurs only in $D\geq 5$ and it is the massless $[2,1^{D-4}]$ field.  In fact, there are no trialities among any of the fields in flat space.

\vspace{20pt}

The examples discussed above illustrate the following general features: dualities can be between two generic massive fields, between two shift-symmetric fields, between two PM fields, or between a PM and a shift symmetric field.  When the duality is between two massive fields, the duality relation takes the form of a mapping between one field and the derivative of its dual, of the schematic form 
\be ({\rm field})\sim \epsilon\, \nabla\, ({\rm dual\ field}), \ \ \ ({\rm  dual\  field})\sim {1\over m^2}\epsilon\, \nabla\, ({\rm field}),\ee
where $m$ is a mass scale related to the masses of the fields, and the consistency of the two equations holds on-shell.  The masses of the two fields on either side of the duality are related such that their dual conformal dimensions computed from \eqref{mixsymmassde2ree} are equal.  When the duality involves PM/shift-symmetric fields, then it takes the form of a mapping between field strengths, either a direct algebraic mapping of the form
\be ({\rm field\ strength})\sim \epsilon\,  ({\rm dual\ field\ strength})\,, \ee
one involving a derivative as in the massive case,
\be ({\rm field\ strength})\sim \epsilon\, \nabla\, ({\rm dual\ field\ strength}), \ \ \ ({\rm  dual\  field\ strength})\sim {1\over m^2}\epsilon\, \nabla\, ({\rm field\ strength})\, ,\ee
or, in some cases, a direct identification of field strengths, with no epsilon symbol,
\be ({\rm field\ strength})\sim   ({\rm dual\ field\ strength})\,. \ee 
In the sections that follow, we give a listing of all the various dualities that occur among the fields of all types in all dimensions.

\section{Dualities on (A)dS\label{adstabesecees}}

We now turn to cataloging all the dualities that arise among massive, PM and shift-symmetric fields on (A)dS$_D$.  All the dualities listed below can be checked using their Weyl modules, and the basic duality relations between the fields or field strengths can be found by looking at how they match up within the dual Weyl modules, in the manner illustrated in Section \eqref{secweylmoddualsec} above.  

In any given dimension $D$, for massive fields, only certain tableaux give propagating degrees of freedom, and they all have $\leq D-1$ rows.  Of these, there are those with $\geq D/2$ rows that we call non-minimal.  All of these dualize into a massive field with a tableau that has $<D/2$ rows.  For each of these dual pairs, the non-minimal tensor has more indices in the first column of its tableau than the other.  We call the shorter tableau the minimal tensor (mirroring the terminology for the orthogonal group tensor representations reviewed in Appendix \ref{sodrepappendix}).  All the PM points and shift symmetric points of the non-minimal fields get related to PM and shift-symmetric points of minimal fields.

As illustrated above, for fields that are self-dual, the field can be split into irreducible chiral halves.  The self-duality can either be imaginary or real according to whether the basic duality relations between the fields involve $i$'s or not.  When the self-duality is real, there should exist a free (A)dS invariant Lagrangian theory that propagates the two chiral halves independently; in the massive cases this should be accomplished through the addition of Chern-Simons-like terms, whereas in the PM cases they may be complications analogous to those encountered for the chiral $p$-forms.  In those cases where the self-duality is imaginary, the two irreducible components must both be present in identically propagating pairs; this should be reflected by the non-existence of real-valued, single-field Chern-Simons-like terms for the massive fields, and the non-existence of single-field chiral actions for PM cases.  

In what follows, when tabulating the dualities, when we write a tableau with a certain shape, it means that all the rows that are shown are non-zero.  Thus, e.g. $[s]$  means all one row tableaux with $s\geq 1$ and $[s_1,s_2,s_3]$ means all three row tableaux, with $s_1\geq s_2\geq s_3\geq 1$.  Dualities will be indicated with arrows: a solid arrow $\dualarrow$ indicates a duality where the basic duality relation involves the epsilon symbol, a hollow arrow $\dualarrowline$  indicates a duality where the basic duality relation does not involve the epsilon symbol.  

\subsection{Massive dualities on (A)dS}

Fields of generic mass on (A)dS have the same degree of freedom content as their flat space counterparts; there are no discontinuities as the background curvature tends to zero with the mass held fixed.  Thus the set of massive fields which carry dynamical degrees of freedom\footnote{It is expected that a field which propagates no local degrees of freedom should have no states, or at most a finite number of states, in its Weyl module.  But in some cases, simply removing the non-permissible tableaux does not always leave a finite number of states.  For example, a massive spin-2 field in $D=2$ is non-dynamical yet its Weyl module, once all the non-permissible states are removed and the rest dualized to their minimal forms, still has the infinite set of local operators $[l]$, $l\geq 2$, consisting of the row built from the field itself.  The same is true for many other examples, including the massive field $[s,1^{D-2}]$ with $s\geq 2$ in $D$ dimensions whose Weyl modules consists of the states $[l]$ with $l\geq s$.  In these cases, there should be additional equation of motion constraints that remove the extra states, and the correlation functions of all local operators should be trivial in the sense that they are purely contact terms.  This can be seen for example with the massive spin-2 in flat space; its momentum space propagator (found for example in eq. 2.44 of \cite{Hinterbichler:2011tt}), whether we take the traceless part of the field or not, does not vanish in $D=2$ but reduces to a constant, corresponding to a delta function contact term in position space.  This occurs because the kinetic term is a total derivative in $D=2$ and so only the Fierz-Pauli mass term contributes to the equations of motion which then yields the constraint $h_{\mu\nu}=0$, removing the remaining local operators of the Weyl module.} and the pattern of massive dualities among them is exactly the same as the flat space version detailed in Appendix \eqref{massivedflatsection}, and there is no need to repeat it here.   In each case, we can also confirm that they work out using their Weyl modules.  

Unlike flat space, on (A)dS the bare field masses $\tilde m$ on each side of the duality are not equal, but are instead related such that their dual CFT conformal dimensions computed from \eqref{mixsymmassde2ree} are equal.    This reflects the equivalence of the underlying (A)dS representations, realized for example through their dual CFT Verma modules, where the primary states are dual pairs under $SO(D-1)$ (the same as the flat space little group).

\subsection{Shift symmetric dualities on (A)dS }

The massive dualities link together the shift symmetric fields.  The pattern by which this happens is as follows:
 \begin{itemize}

 \item{\framebox{$D=2$:}} The shift symmetric scalars and vectors are dual to each other, and they are also dual to themselves:
 \be \overset{\dualarrowcurved}{ \underset{0}{[0]}},  \overset{\dualarrowcurved}{\underset{k+1}{[0]}}\dualarrow \overset{\dualarrowcurved}{\underset{ k}{[1]}},\  \ee
 where $k=0,1,2,\ldots$.
The self-dualities are real. 

 \item{\framebox{$D\geq 4$ even:}}  Let $D=2p+2$, $p=1,2,\ldots$.  We have the following duality equivalences among the massive shift symmetric fields,
 \bea && \underset{k+1}{ [0]} \dualarrow  \underset{k}{[1^{D-1}]} ,\ \ \underset{k}{ [s_1]}\dualarrow \underset{k}{[s_1,1^{D-3}]} ,\ \ \underset{k}{[s_1,s_2]}\dualarrow\underset{k}{ [s_1,s_2,1^{D-5}]} ,\ldots, \nn\\
&&  \underset{k}{ [s_1,\ldots,s_p] } \dualarrow \underset{k}{[s_1,\ldots,s_p,1] }\,, \label{Doddsequivele}\eea
where $k=0,1,2,\ldots$.
 
 \item{\framebox{$D$ odd:}}   Let $D=2p+1$, $p=1,2,\ldots$.   We have the following duality equivalences among the massive shift symmetric fields:
 \bea && \underset{k+1}{[0]}\dualarrow  \underset{k}{[1^{D-1}]},\ \ \underset{k}{[s_1]}\dualarrow \underset{k}{[s_1,1^{D-3}]} ,\ \ \underset{k}{[s_1,s_2]}\dualarrow \underset{k}{ [s_1,s_2,1^{D-5}] },\ldots, \nn\\
&& \underset{k}{ [s_1,\ldots,s_{p-1}]}  \dualarrow \underset{k}{[s_1,\ldots,s_p,1,1] }\,,\label{firstdoddeede}\eea
where $k=0,1,2,\ldots$.

The $p$-row shift symmetric fields are dual to themselves:  
\be \underset{k}{\overset{\dualarrowcurved}{[s_1,\ldots,s_p]}}\, , \ee
and thus split into chiral halves $\underset{k}{{[s_1,\ldots,s_p]_\pm}}$.
 For $p$ even ($D=5,9,13,\ldots$) the self-duality relations are imaginary, for $p$ odd ($D=3,7,11,\ldots$) the self-duality  relations are real.

\end{itemize}

\subsection{PM/shift symmetric dualities on (A)dS\label{PMshsdualsection}}

There are several cases in which dualities link PM fields with shift symmetric fields:  

 \begin{itemize}
 
 \item{\framebox{$D=3$:}} The maximal depth PM fields are dual to shift symmetric scalars,
 \be  \underset{k}{[0]}\dualarrow \underset{ \ \nabla^{k+1}}{[k+1]}\,,\ee
 where $k=0,1,2,\ldots$.  Due to the first duality of \eqref{firstdoddeede}, those with top row length $\geq 2$ are also dual to the shift symmetric 2-forms,
 \be  \underset{k}{[1,1]}\dualarrowline \underset{ \ \nabla^{k+2}}{[k+2]}\,,\ee
 where $k=0,1,2,\ldots.$  (This duality is indicated with a hollow arrow because it does not involve the epsilon symbol; the fields strengths on the two sides are directly equated.)

 \item{\framebox{$D\geq 4$:}} 
A PM long hook in which the bottom square is activated is dual to a shift symmetric scalar,
\be \underset{k}{[0]}\dualarrow \underset{\ \ \ \ \ \ \ \ \   \nabla}{[k+1,1^{D-3}]}\,.\ee
where $k=0,1,2,3,\ldots$.  Due to the first duality of \eqref{Doddsequivele}, those with top row length $\geq 2$ are also dual to the shift symmetric $(D-1)$-forms,
 \be  \underset{k}{[1^{D-1}]}\dualarrowline \underset{\ \ \ \ \ \ \ \ \nabla}{[k+2,1^{D-3}]}\,,\ee
 where $k=0,1,2,\ldots.$  

 \end{itemize}

\subsection{PM dualities on (A)dS }

Here are the dualities that link PM fields with other PM fields:  

 \begin{itemize}
 
 \item{\framebox{$D$ even:}}   Let $D=2p+2$, $p=1,2,\ldots$.   First consider the PM cases among the non-minimal tensors where anything except for the bottom square is activated.   In these cases, they are dual to the corresponding minimal tensors with the same blocks activated
\bea && \hspace{-20pt} [\underset{\ \nabla^{\cdots}}{s_1},1^{D-3}] \dualarrow [\underset{\ \nabla^{\cdots}}{s_1}]\ , \ \  [\underbrace{s_1,s_2}_{\nabla^{\cdots}},1^{D-5}] \dualarrow  [\underbrace{s_1,s_2}_{\nabla^{\cdots}}] \ ,\ \ldots\ ,\  [\underbrace{s_1,\ldots,s_{p-1}}_{\nabla^{\cdots}} ,1,1,1]  \dualarrow  [\underbrace{s_1,\ldots,s_{p-1}}_{\nabla^{\cdots}}] \, , \nn\\ 
&& \hspace{-20pt}  [\underbrace{s_1,\ldots,s_{p-1}}_{\nabla^{\cdots}} ,s_p,1]  \dualarrow  [\underbrace{s_1,\ldots,s_{p-1}}_{\nabla^{\cdots}},s_p] \,,  \    \overset{\ \ \dualarrowcurved}{  [s_1,\ldots,s_{p-1},\underset{\ \nabla^{\cdots}}{s_p},1] }\dualarrow   \overset{\ \ \dualarrowcurved}{  [s_1,\ldots,s_{p-1},\underset{\ \nabla^{\cdots}}{s_p}] }\ .\label{Dodddequivele}
\eea
Here $\nabla^{\cdots}$ indicates that any of the indicated squares on the left hand side with the indicated $s$'s can be activated, and the same squares are activated on the right hand side.  In the $p$-row case with all of the last row activated we have the self-duality
\be  \hspace{-22pt}  \overset{\dualarrowcurved}{ \underset{\ \ \ \ \ \ \ \ \ \ \ \ \  \ \ \ \ \ \nabla^{s_p}}{ [s_1,\ldots,s_{p-1},s_p] }}\, .\label{Dodddequivel2e} \ee
The self-dualities in these expressions are real for $p$ even ($D=6,10,14,\ldots$) and imaginary for $p$ odd ($D=4,8,12,\ldots$).

Now consider the non-minimal tensors where the bottom block is activated.  We have the following pattern of dualities:
\bea && \underset{\ \ \ \ \ \ \ \ \nabla}{  [s_1,s_2,1^{D-5}]}\   \dualarrow \hspace{-20pt}  \underset{\ \ \ \ \ \ \ \nabla^{s_1-s_2+1}}{[s_1]}  ,\ \ \  \underset{\ \ \ \ \ \ \ \ \ \ \nabla}{  [s_1,s_2,s_3,1^{D-7}]}\ \dualarrow \hspace{-21pt}  \underset{\ \ \ \ \ \ \ \ \ \ \ \   \nabla^{s_2-s_3+1}}{[s_1,s_2]} , \nn\\ 
&&   \underset{\ \ \ \ \ \ \ \ \ \ \ \  \ \ \ \  \nabla}{  [s_1,s_2,s_3,s_4,1^{D-9}]} \ \dualarrow\hspace{-21pt}   \underset{\ \ \ \ \ \ \ \ \ \ \ \ \ \ \ \ \nabla^{s_3-s_4+1}}{[s_1,s_2,s_3]} \,, \ \ldots\, , \ \ \nn\\
&&  \underset{\ \ \ \ \ \ \ \ \ \ \ \ \ \nabla}{  [s_1,\ldots,s_p,1]} \ \dualarrow  \hspace{-27pt}  \underset{\ \ \ \ \ \  \ \ \ \ \ \ \ \ \ \ \ \ \ \ \ \nabla^{s_{p-1}-s_p+1}}{[s_1,\ldots,s_{p-1}]} \, .
\eea
(The case $\underset{\nabla}{[1^{D-1}]}$ is non-dynamical, as covered in Section \ref{nondynsection}, and the case $\underset{\ \ \ \ \nabla}{[s_1,1^{D-3}]}$ is dual to the shift symmetric scalars, as covered in Section \ref{PMshsdualsection}.)

Due to \eqref{Dodddequivele}, we also have the following direct relations:
\bea  &&\underset{\hspace{33pt}\nabla}{ [s_1,s_2,1^{D-5}]}\ \dualarrowline\underset{\hspace{-2pt}\nabla^{s_1-s_2+1}}{ [s_1,1^{D-3}] }\,\  ,\  \underset{\hspace{50pt}\nabla}{ [s_1,s_2,s_3,1^{D-7}]}\ \dualarrowline\underset{\hspace{18pt}\nabla^{s_2-s_3+1}}{ [s_1,s_2,1^{D-5}] }\, \ ,\ \ldots\ , \nn\\
&& \underset{\hspace{52pt}\nabla}{ [s_1,\ldots,s_p,1]}\ \dualarrowline\underset{\hspace{43pt}\nabla^{s_{p-1}-s_p+1}}{ [s_1,\ldots,s_{p-1},1,1,1] }\,.
\eea
(Here the values for the $s$'s must be consistent with the shape and existence of the indicated PM points, i.e. $s_1\geq 2$ for the first relation, $s_2\geq 2$ for the second, $\ldots\ $, $s_{p-1}\geq 2$ for the last.)

 \item{\framebox{$D$ odd:}}   Let $D=2p+1$, $p=2,3,\ldots$.   First consider the PM cases among the non-minimal tensors where anything except for the bottom square is activated.   In these cases, they are dual to the corresponding minimal tensors with the same blocks activated,
\bea && \hspace{-20pt} [\underset{\ \nabla^{\cdots}}{s_1},1^{D-3}] \dualarrow [\underset{\ \nabla^{\cdots}}{s_1}]\ , \ \  [\underbrace{s_1,s_2}_{\nabla^{\cdots}},1^{D-5}] \dualarrow  [\underbrace{s_1,s_2}_{\nabla^{\cdots}}] \ ,\ \ldots\ ,\  [\underbrace{s_1,\ldots,s_{p-1}}_{\nabla^{\cdots}} ,1,1]  \dualarrow  [\underbrace{s_1,\ldots,s_{p-1}}_{\nabla^{\cdots}}] \, . \nn\\ \label{Dodddequivele232}
\eea
Here $\nabla^{\cdots}$ means that any of the indicated squares on the left hand side can be activated, with the same squares activated on the right hand side.

There are two special cases involving the $p$ row tableaux: self-duality if any squares not in the bottom row are activated,
\be \overset{\ \ \ \ \ \  \dualarrowcurved}{ \ \ \ \ \ \  [\underbrace{s_1,\ldots,s_{p-1}}_{\nabla^{\cdots}},s_{p}] }\, , \ee
which is imaginary if $p$ is even ($N=5,9,13,\ldots$) and real if $p$ is odd ($N=7,11,\ldots$),
and a duality if all the squares in the $p$-th row are activated,
\be \underset{\ \ \ \ \ \ \ \ \ \ \ \ \ \ \ \ \ \ \nabla^{s_p}}{  [s_1,\ldots,s_{p-1},s_{p}] } \ \dualarrow   \hspace{-25pt}  \underset{\ \ \ \ \ \ \ \ \ \ \ \ \ \ \ \ \ \ \ \ \ \ \ \ \nabla^{s_{p-1}-s_p+1}}{  [s_1,s_2,\ldots,s_{p-1}] } .\ee

Now consider the non-minimal tensors where the bottom block is activated.  We have the following pattern of dualities:
\bea && \underset{\ \ \ \ \ \ \ \ \nabla}{  [s_1,s_2,1^{D-5}]} \ \dualarrow\hspace{-20pt}   \underset{\ \ \ \ \ \ \ \nabla^{s_1-s_2+1}}{[s_1]}  ,\ \ \  \underset{\ \ \ \ \ \ \ \  \ \ \ \ \nabla}{  [s_1,s_2,s_3,1^{D-7}]}\ \ \dualarrow \hspace{-20pt}  \underset{\ \ \ \ \ \ \ \ \ \ \ \   \nabla^{s_2-s_3+1}}{[s_1,s_2]} , \nn\\ 
&&   \underset{\ \ \ \ \ \ \ \ \ \ \ \ \ \ \ \ \nabla}{  [s_1,s_2,s_3,s_4,1^{D-9}]}\ \ \dualarrow  \hspace{-20pt}  \underset{\ \ \ \ \ \ \ \ \ \ \ \ \ \ \ \ \nabla^{s_3-s_4+1}}{[s_1,s_2,s_3]} \,, \ \ldots\ , \ \ \nn\\
&&  \underset{\ \ \ \ \ \ \ \ \ \ \ \ \ \ \  \ \ \ \ \nabla}{  [s_1,\ldots,s_{p-1},1,1]} \ \ \dualarrow \hspace{-25pt}  \underset{\ \ \ \ \ \ \ \ \ \ \ \ \ \ \ \ \ \ \ \ \ \ \ \ \ \ \ \nabla^{s_{p-2}-s_{p-1}+1}}{[s_1,\ldots,s_{p-3},s_{p-2}]} \,.
\eea
(The case $[1^{D-1}]$ is non-dynamical, as covered in Section \ref{nondynsection}, and the case $[s_1,1^{D-3}]$ is dual to the shift symmetric scalars, as covered in Section \ref{PMshsdualsection}.)

We  have the following direct relations:
\bea  &&\underset{\hspace{33pt}\nabla}{ [s_1,s_2,1^{D-5}]}\ \dualarrowline\ \underset{\hspace{-2pt}\nabla^{s_1-s_2+1}}{ [s_1,1^{D-3}] }\,\  ,\  \underset{\hspace{50pt}\nabla}{ [s_1,s_2,s_3,1^{D-7}]}\ \dualarrowline\ \underset{\hspace{18pt}\nabla^{s_2-s_3+1}}{ [s_1,s_2,1^{D-5}] }\, \ ,\ \ldots\ , \nn\\
&& \underset{\hspace{73pt}\nabla}{ [s_1,\ldots,s_{p-1},1,1]}\ \dualarrowline\ \underset{\hspace{43pt}\nabla^{s_{p-2}-s_{p-1}+1}}{ [s_1,\ldots,s_{p-2},1,1,1,1] }\,, \nn\\
&& \hspace{-3pt} \underset{\hspace{45pt}\nabla^{s_p}}{ [s_1,\ldots,s_{p}]}\ \dualarrowline\ \underset{\hspace{53pt}\nabla^{s_{p-1}-s_{p}+1}}{ [s_1,\ldots,s_{p-1},1,1] }\,.
\eea
(Here the values for the $s$'s must be consistent with the shape and existence of the indicated PM points, i.e. $s_1\geq 2$ for the first relation, $s_2\geq 2$ for the second, $\ldots\ $, $s_{p-1}\geq 2$ for the last.)

 \end{itemize}

\subsection{Non-dynamical PM fields\label{nondynsection}}

There are some PM points where the field becomes non-dynamical, i.e the corresponding massive field has propagating degrees of freedom but the PM point does not.  A simple example is the spin-1 field in $D=2$: the massive photon has one propagating degree of freedom but the massless photon has none.  These non-dynamical PM fields account for the PM points which do not participate in the above dualities; once these are taken into account, all the possible PM points for non-minimal tensors are related to PM or shift symmetric points of minimal tensors.

 \begin{itemize}
 
 \item{\framebox{$D$ even:}}  The only massive field which has an empty PM limit is the $D-1$ form field,
 \be \underset{\nabla}{[1^{D-1}]}\, .\ee
 This has only one PM point where the bottom square is activated, becoming a massless $D-1$ form which has no propagating degrees of freedom.
 
  \item{\framebox{$D$ odd:}}  Let $D=2p+1$, $p=1,2,3,\ldots$.  The $p$-row fields become non-dynamical when any number less than all the squares in the bottom row are activated (it remains dynamical if all $s_p$ of the bottom rows are activated),
  \be\underset{\ \ \ \ \ \ \ \  \  \ \nabla^t}{ [s_1,\ldots,s_p]}\, ,\ee
  where $t<s_p$.

 In addition, the $D-1$ form field,
 \be \underset{\nabla}{[1^{D-1}]}\, ,\ee
  has only one PM point where the bottom square is activated, becoming a massless $D-1$ form which has no propagating degrees of freedom.

 \end{itemize}

\section{Summary of (A)dS dualities in low dimensions}

Here, for ease of reference, we go through the lower dimensional cases explicitly.  

 \begin{itemize}
 
 \item{\framebox{$D=2$:}} 
 \textbf{Massive fields:} The only massive fields with non-trivial dynamics are the scalar and vector,
 \be [0],\ \ [1] .\ee
 \textbf{Massive dualities:}  They are dual to each other,
  \be [0]\dualarrow [1] .\ee
 Both carry the same single degree of freedom; the vector is the non-minimal tensor and the scalar the minimal one.
 
 \textbf{Non-dynamical PM points:} 
 The only PM point among the dynamical massive fields is the massless vector, 
 \be \underset{\nabla}{[1]},\ee
which carries no degrees of freedom.
 
   \textbf{PM and shift symmetric dualities:}  
 The shift symmetric scalars and vectors are dual to each other, and they are also dual to themselves:
 \be \overset{\overset{{\rm Re}}{ \dualarrowcurved}}{ \underset{0}{[0]}},\ \ \  \overset{\overset{{\rm Re}}{ \dualarrowcurved}}{\underset{k+1}{[0]}}\dualarrow \overset{\overset{{\rm Re}}{ \dualarrowcurved}}{\underset{ k}{[1]}}\, ,\ee
 where $k=0,1,2,\ldots$.
The self-dualities are real, and the fields split into chiral parts $\underset{\hspace{-7pt}k}{[0]_\pm}$, $\underset{\hspace{-7pt}k}{[1]_\pm}$.

  \item{\framebox{$D=3$:}}   \textbf{Massive fields:} The only massive fields with propagating degrees of freedom are the scalar, symmetric tensors and 2-form,
 \be [0],\ \ [s]\, ,\ \ [1,1]\, .\ee

 \textbf{Massive dualities:} The scalar and 2-form are dual to each other, and each carries one degree of freedom,
\be  [0]\dualarrow [1,1]\,.\ee

The symmetric tensor fields $[s]$ all have 2 degrees of freedom and are dual to themselves with real duality relations, 
\be \overset{\overset{{\rm Re}}{ \dualarrowcurved}}{ [s]} \,.\ee
They can be split into chiral halves, $ [s]_\pm$, each of which carries one degree of freedom.

 \textbf{Non-dynamical PM points:} 
Among the PM points, the symmetric tensors with less than maximal depth, and the massless 2-form, are non-dynamical,
\be \underset{\nabla^t}{[s]}\, ,\ \ \  \underset{\ \ \ \nabla}{[1,1]}\,,\ee 
where $t=1,2,\ldots, s-1$.  The case $t=1$ for $s\geq 2$ is the well-known statement that gravity and the massless higher spins are topological in $D=3$.

   \textbf{PM and shift symmetric dualities:}  
Among the shift symmetric and PM fields, we have the the following dualities:
\be \raisebox{-35pt}{\epsfig{file=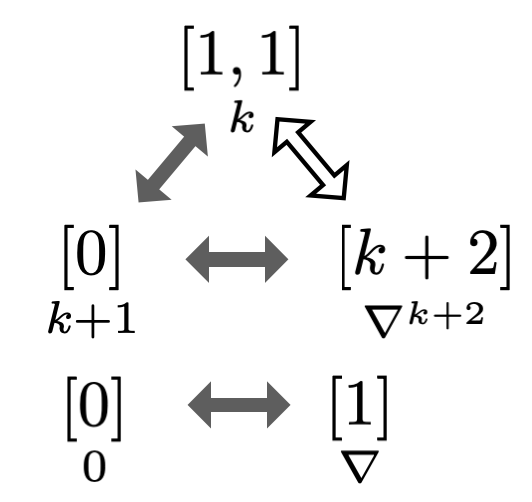,height=1.3in,width=1.3in}}\,  , \ \ \underset{k}{ \overset{\overset{{\rm Re}}{ \dualarrowcurved}}{ [s]} }\,,  \ee  
where $k=0,1,2,\ldots$.  The self-dual shift symmetric fields split into real chiral parts $\underset{\hspace{-8pt} k}{ [s]_{\pm}}$.

   \item{\framebox{$D=4$:}}  \textbf{Massive fields:} The only massive fields with propagating degrees of freedom are 
 \be [0],\ \ [s]\, ,\ \ [s,1]\,, \ \ [1,1,1]\, .\ee

 \textbf{Massive dualities:}
The dualities between them are
 \be [0]\dualarrow [1,1,1]\, ,\ \ \  [s]\dualarrow [s,1]\, .\ee

 \textbf{Non-dynamical PM points:} 
  Among the PM points, the only one which is non-dynamical is the massless 3 form,
  \be \underset{\ \ \ \ \ \ \nabla}{[1,1,1]}\, .\ee
   
   \textbf{PM and shift symmetric dualities:}  
  Among the shift symmetric and PM fields, we have the dualities
\bea \raisebox{-35pt}{\epsfig{file=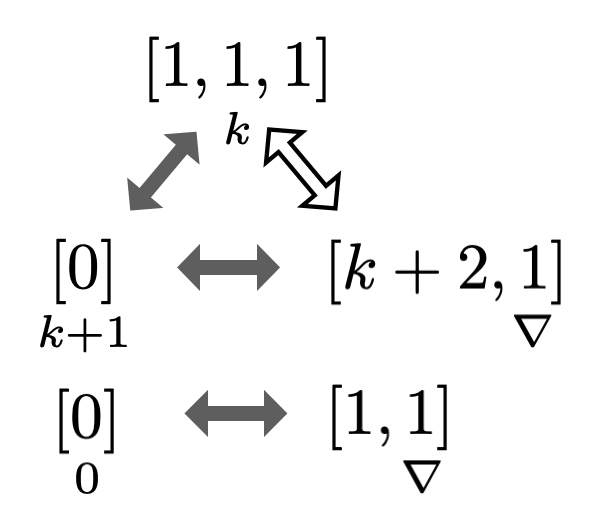,height=1.3in,width=1.4in}}\, ,\ \ \ \  \underset{k}{[s]} \dualarrow \underset{k}{[s,1] }\ , \eea
where $k=0,1,2,\ldots$, and
 \be  \underset{\nabla^t}{\ \  \overset{\overset{{\rm Im}}{ \dualarrowcurved}}{ [s,1]} }\ \dualarrow \underset{\nabla^t}{ \overset{\overset{{\rm Im}}{ \dualarrowcurved}}{ [s]} }\,, \underset{\nabla^s}{ \overset{\overset{{\rm Im}}{ \dualarrowcurved}}{ [s]} } \,, \ee 
   where $t=1,2,\ldots,s-1$.  In particular, all the symmetric tensor PM fields are imaginary dual to themselves \cite{Deser:2013xb,Hinterbichler:2016fgl}, and the PM hook with the top block activated is dual to the massless graviton \cite{Basile:2015jjd}.
   
    \item{\framebox{$D=5$:}} \textbf{Massive fields:} The only massive fields with propagating degrees of freedom are 
 \be [0],\ \ [s_1]\, ,\ \ [s_1,s_2]\, ,\ \ [s_1,1,1]\,, \ \  [1,1,1,1]\, .\ee

 \textbf{Massive dualities:}
The dualities between them are
 \be [0]\dualarrow [1,1,1,1]\, ,\ \ \  [s_1]\dualarrow [s_1,1,1]\, ,\ \ \ \overset{\overset{{\rm Im}}{ \dualarrowcurved}}{[s_1,s_2]}\,.\ee
The self-duality of the two-row fields is imaginary

 \textbf{Non-dynamical PM points:} 
  Among the PM points, the only ones which are non-dynamical are the massless 4 form and the 2-row PM fields with any number of squares $<s_2$ in the bottom row activated,
  \be \underset{\ \ \ \ \ \ \ \ \  \nabla}{[1,1,1,1]}\, ,\ \ \ \underset{\ \ \ \ \ \nabla^t}{[s_1,s_2]}\, , \ee
   where $t=1,2,\ldots,s_2-1$.
  
   \textbf{PM and shift symmetric dualities:}  
     Among the shift symmetric and PM fields, we have the dualities
\bea \raisebox{-35pt}{\epsfig{file=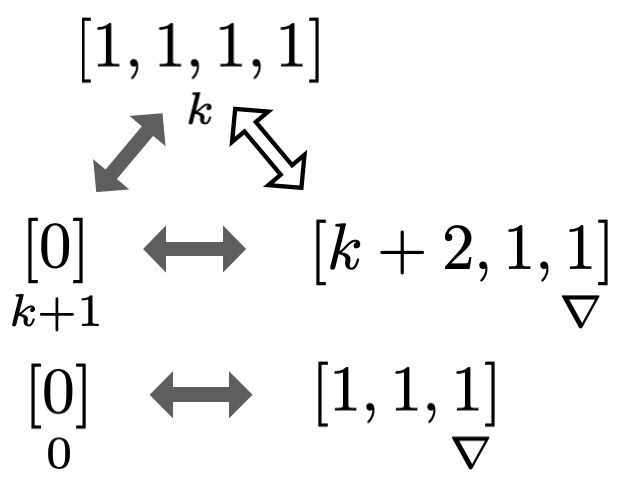,height=1.3in,width=1.5in}}\, ,\ \ \ \  \underset{k}{[s_1]} \dualarrow \underset{k}{[s_1,1,1] }\ , \ \ \ \underset{k}{\overset{\overset{{\rm Im}}{ \dualarrowcurved}}{[s_1,s_2]}} \, ,\eea
where $k=0,1,2,\ldots$, and
\be  \raisebox{-35pt}{\epsfig{file=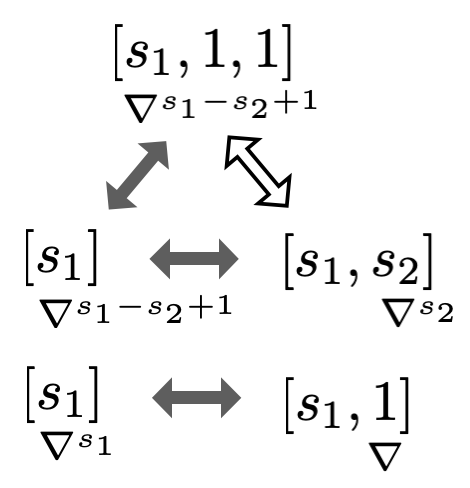,height=1.5in,width=1.3in}}\,  ,\ \ \  \ \ \  \ \underset{\hspace{-15pt} \nabla^t}{ \overset{\overset{{\rm Im}}{ \dualarrowcurved}}{ [s_1,s_2]} } \,, \ee 
   where $s_2\geq 2$ and $t=1,2,\ldots ,s_1-s_2$.

     \item{\framebox{$D=6$:}}  \textbf{Massive fields:} The only massive fields with propagating degrees of freedom are 
 \be [0],\ \ [s_1]\, ,\ \ [s_1,s_2]\, ,\ \ [s_1,s_2,1]\,, \ \ [s_1,1,1,1]\,, \ \  [1,1,1,1,1]\, .\ee

 \textbf{Massive dualities:}
The dualities between them are
 \be [0]\dualarrow [1,1,1,1,1]\, ,\ \ \  [s_1]\dualarrow [s_1,1,1,1]\, ,\ \ \ [s_1,s_2]\dualarrow [s_1,s_2,1]\, .\ee

 \textbf{Non-dynamical PM points:} 
  Among the PM points, the only one which is non-dynamical is the massless 5 form,
  \be \underset{\ \ \ \ \ \ \ \ \ \ \ \nabla}{[1,1,1,1,1]}\, . \ee

   \textbf{PM and shift symmetric dualities:}  
   
      Among the shift symmetric and PM fields, we have the dualities
\bea \raisebox{-40pt}{\epsfig{file=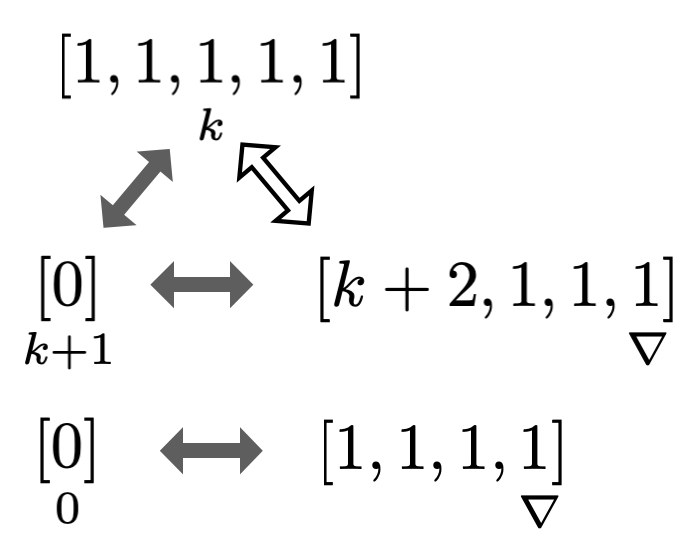,height=1.4in,width=1.8in}}\, ,\ \ \ \  \underset{k}{[s_1]} \dualarrow \underset{k}{[s_1,1,1,1] }\ ,\ \ \ \underset{k}{[s_1,s_2]} \dualarrow \underset{k}{[s_1,s_2,1] } \, \ \ ,\eea
where $k=0,1,2,\ldots$, 
\be  \raisebox{-40pt}{\epsfig{file=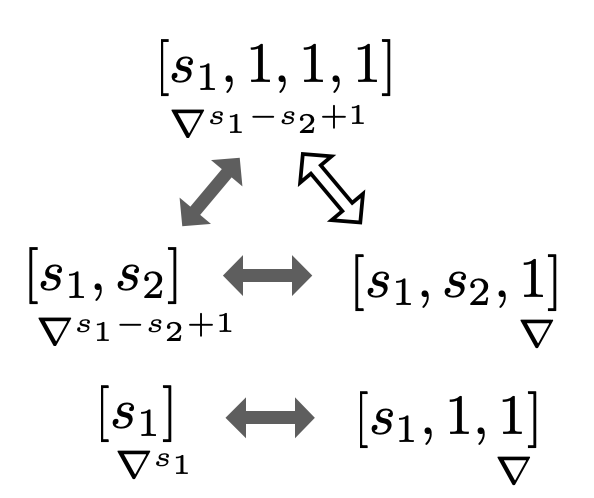,height=1.4in,width=1.6in}}\,\ \  ,\ee
where $s_2\geq 2$, and
\bea  
&& \underset{\hspace{-15pt}\nabla^{t} }{ [s_1,s_2] } \dualarrow \underset{\hspace{-25pt}\nabla^{t} }{ [s_1,s_2,1] } \, ,\ \ \underset{\hspace{20pt} \nabla^{t} }{{ \overset{\overset{{\rm Re}}{ \dualarrowcurved}}{ [s_1,s_2]} }} \dualarrow \underset{\hspace{10pt}\nabla^{t} }{{ \overset{\overset{{\rm Re}}{ \dualarrowcurved}}{ [s_1,s_2,1]} }} \, , \ \   \underset{\hspace{24pt} \nabla^{s_2}}{ \overset{\overset{{\rm Re}}{ \dualarrowcurved}}{ [s_1,s_2]} } \,,  \eea 
   where $t=1,2,\ldots,s_1-s_2$ in the first relation and  $t=1,2,\ldots,s_2-1$ in the second.

  \end{itemize}

\section{Conclusions}

We have cataloged the electromagnetic-type dualities that exist between canonical fields on (A)dS$_D$, for all $D\geq 2$.  Fields on (A)dS can be massive, partially massless, or shift symmetric; massive fields are generic, and the PM and shift symmetric fields occur at specific mass values relative to the background curvature scale.  The dualities can occur between two massive fields, between two shift symmetric fields, between a shift symmetric and a PM field, or between two PM fields.  In certain cases there are also self-dualities in a field of any of these types.
When there are dualities between two different field types, it means that the same dynamical degrees of freedom, transforming as some irreducible representation of the background isometry group, can be packaged in two different canonical ways.  The dualities relate all PM and shift symmetric fields with non-minimal tableaux, i.e. those with $\geq D/2$ rows, to those with $<D/2$ rows.  In addition there are some trialities, a phenomenon that does not occur in flat space.

Here we tabulated only the dualities among the bosonic fields, but a similar classification will also hold for the fermionic fields, which possess PM and shift symmetric points analogous to their bosonic counterparts \cite{Bonifacio:2023prb}.  Self-duality of the spin $s$ PM fermions in $D=4$ was studied in \cite{Deser:2014ssa}.

Just as on flat space the dualities reflect the equivalence of the underlying Poincare representations, on AdS space these dualities reflect the underlying equivalence between the AdS representations that the fields propagate, as realized e.g. through their dual CFT Verma modules.  This is reflected in the identification of the dual CFT conformal dimension  \eqref{mixsymmassde2ree} across dual pairs, and in the equivalence of the two $SO(D-1)$ representations of the dual CFT primary states (which is the same irreducible representation as that of the massive little group on flat space).  On dS space, the dualities reflect the equivalence between their underlying dS representations, as classified in e.g. \cite{10.1063/1.1665471}.

On dS space, these dualities are important to take into account in understanding the map between PM/shift symmetric fields and the classification of dS representations \cite{Basile:2016aen}.  For example, in $D=3$ we are told there are no discrete series representations \cite{Bargmann:1946me,zbMATH03056355} (see Section 2.3 of \cite{Bonifacio:2023ban} for a concise review).  But this presents a puzzle because the PM and shift-symmetric fields are supposed to live in discrete series representations and there are spin $s$ PM fields and shift-symmetric scalars which are unitary on dS$_3$ and hence should be accounted for in the list of dS irreducible representations.  The resolution is provided by duality and by the triviality of most of the PM points: only the maximal depth PM points are dynamical, but these points are at $\Delta=1$, so this single discrete series point intersects with the principal series and is included there.  Furthermore, these maximal depth PM points are dual to the shift symmetric scalars, so the discrete series for the scalars are indeed present, but they do not count as distinct irreducible representations since they are already accounted for by the spinning $\Delta=1$ points.

In some cases these dualities must also be accounted for when assessing the unitarity of partially massless or shift symmetric points when realized using their non-minimal tensors.  The generic rule for the unitarity of PM fields is that they are unitary on dS only if any number of squares on the bottom block are activated, and unitary on AdS only if a single square on the top block is activated. The generic rule for shift-symmetric fields is that only the scalars are unitary on dS, and all the shift-symmetric fields are unitary on AdS.  An example which breaks these rules is the dualities between level $k+1$ shift symmetric scalars and the level $k$ $(D-1)$-forms, which are both unitary on both AdS and dS.  Another is the $\underset{\hspace{-28pt}\nabla^t}{ [s,1^{D-3}]}$ for $1\leq t \leq s-1$ which is unitary on dS because it is dual to $\underset{\nabla^t}{[s]}$ which is unitary on dS.  These examples evade the generic unitarity arguments because some of the component degrees of freedom which would have had wrong sign kinetic terms end up being non-propagating.  There could still be issues with unitarity if some of these non-unitary non-propagating degrees of freedom still have residual edge modes or other topological effects that cannot be ignored.   (This same mechanism allows certain higher-derivative theories to be unitary in certain dimensions \cite{Joung:2016naf}, the best known example being new massive gravity in $D=3$ \cite{Bergshoeff:2009hq}.)

For fields that are dual to themselves, duality is a symmetry of the free theory, and a natural question is whether there are interactions that preserve this duality symmetry.  It is by no means guaranteed that an interacting theory, even if uniquely interesting, preserves the linear duality symmetry.  For example, the massless spin-2 theory in flat $D=4$ is dual to itself, but its unique 2-derivative non-abelian extension, GR, does not preserve this duality \cite{Monteiro:2023dev}.  

The simplest interactions that involve the PM or shift symmetric fields are those of the Born-Infeld type.  These are interactions which are constructed from the invariant field strength of the linear theory, and so do not deform the algebra of gauge or shift symmetries.
A classic example of a non-linear theory of this type that preserves duality is the namesake Born-Infeld theory of non-linear electrodynamics \cite{Born:1933qff,Born:1933lls,Born:1933pep,Born:1934gh,Born:1937drv} (its self-duality was apparently first known to Schr\"odinger \cite{Schrodinger:1935oqa}, see also \cite{Gaillard:1981rj,Gibbons:1995cv,Gaillard:1997zr,Avetisyan:2021heg,Sorokin:2021tge}).  Higher form generalizations are also known, see e.g. \cite{Buratti:2019cbm,Bandos:2020hgy,Avetisyan:2022zza}.  Among the PM fields, 
non-linear duality invariant DBI-like equations of motion for maximal depth PM symmetric tensor fields were found in \cite{Cherney:2015jxp}, though it is not known if they derive from covariant Lagrangians.  For the shift-symmetric scalars, certain minimal-derivative DBI-like interactions are known, for example $P(X)$ interactions for the $k=0$ scalar, and (A)dS Galileon interactions for the $k=1$ scalar \cite{Goon:2011qf,Goon:2011uw,Burrage:2011bt,Bonifacio:2021mrf}.  It would be interesting to study whether these or other analogous cases could be used to non-linearly realize the (A)dS dualities.

The most interesting interactions among gauge fields or shift symmetric fields tend to be those that deform the symmetries in a non-trivial way and cannot be constructed from linearized fields strengths.  Some examples for such interaction among the shift symmetric fields are known, such as the $k=1,2$ scalar \cite{Bonifacio:2021mrf} and the $k=0$ vector \cite{Bonifacio:2019hrj}.  It would be interesting to study whether these or other analogous cases could be used as ingredients to non-linearly realize the (A)dS dualities.

Apart from maintaining duality invariance, the dualities should also be kept in mind when looking for interacting theories.  When looking for interactions among the theories with gauge invariance, we can typically expect that the minimal tensor representations allows for more general interactions that are manifestly local at the level of the Lagrangian, whereas the non-minimal tensor may miss some.  For example, if we are interested in interactions of a massless spin-2 particle, if we try to use the dual graviton we would find that we cannot deform its gauge algebra and cannot write 2 derivative interactions \cite{Bekaert:2002uh}.  But if we use the minimal symmetric tensor, we know that we can write general relativity as a consistent 2-derivative theory that deforms the gauge algebra.  The dualities among massive fields should not encounter this problem, and all interactions should be possible on both sides.  However, interactions that are simple in one set of variables will not generally be simple in the dual variables, and restricting to interactions which are simple in both will generally spoil the linearized duality \cite{Hell:2021wzm}.

\vspace{-5pt}
\paragraph{Acknowledgments:}

The author would like to thank James Bonifacio, Karapet Mkrtchyan and Austin Joyce for discussions, and acknowledges support from DOE grant DE-SC0009946.

\appendix

\section{Flat space dualities\label{flatspacedappendix}}

In flat space, we can understand all the dualities through the Poincare representations \cite{Bekaert:2006py} that the dynamical degrees of freedom of the fields carry \cite{Casini:2001gv}.   For a field with spacetime indices in a tableau $[s_1,\ldots, s_p]$, the degrees of freedom are those of a particle transforming under the little group \cite{Weinberg:1995mt} in a representation given by the same tableau.  
Two different fields are dual if their tableaux furnish equivalent representations of the little group, and a field is trivial, i.e. has no propagating degrees of freedom (though it can still be non-trivial topologically), if its little group representation is empty.
 For massive fields in $D$ spacetime dimensions, the little group is $SO(D-1)$ and for massless fields it is $SO(D-2)$.  The irreducible representations of the orthogonal groups and their equivalences are reviewed in appendix \ref{sodrepappendix}; from these we get the various dualities detailed below.  

For each dual pair where the tensors are distinct on each side of the duality, one tensor has more indices in the first column of its tableau than the other.  We call the shorter tableau the minimal tensor, and the longer tableau the non-minimal tensor, mirroring the terminology for the orthogonal group tensor representations reviewed in appendix \ref{sodrepappendix}.  

For fields that are self-dual, the self-duality can either be real or imaginary.  In those case where it is real, CPT does not interchange the two irreducible chiral components, so they can propagate independently in a free Poincare invariant Lagrangian theory.  In those cases where it is imaginary, the two irreducible components are interchanged by CPT and so in the free theory the two chiralities must be present in identically propagating pairs.

\subsection{Massive flat space dualities\label{massivedflatsection}}

 In the massive case, the representations are classified by the massive little group, which for spacetime dimension $D$ is $SO(D-1)$.  Two fields are dual if their tableaux give equivalent representations of $SO(D-1)$, as reviewed in appendix \ref{sodrepappendix}, and if the masses for the two fields are equal.  From this we get the following dualities:
 \begin{itemize}
 
 \item{\framebox{$D$ even:}}  Let $D=2p+2$, $p=0,1,2,\ldots$.  The little group is $SO(2p+1)$.  The only fields with propagating degrees of freedom are those with the following tableaux, corresponding to the permissible tableaux of $SO(2p+1)$, as detailed in appendix \ref{sodrepappendix},
 \bea && [0]\, ,  \ \ [s_1]\, , \  \ [s_1,s_2] \, ,\ \ldots, \ [s_1,\ldots,s_p] \, , \nn\\
&&  [s_1,\ldots,s_p,1]\,  , \ \ldots \ ,  \ [s_1,s_2,1^{D-5}] \, , \ [s_1,1^{D-3}] ,\ \ [1^{D-1}]\,. \label{sodreeprDe} 
\eea
Those in the first row are the minimal tensors and those in the second row are the non-minimal tensors.
 We have the following duality equivalences relating the non-minimal to minimal tensors,
 \bea && [0]\dualarrow  [1^{D-1}]\, ,\ \ [s_1]\dualarrow [s_1,1^{D-3}]\, ,\ \ [s_1,s_2]\dualarrow [s_1,s_2,1^{D-5}] \, ,\ \ldots\ , \nn\\
 && [s_1,\ldots,s_p]  \dualarrow [s_1,\ldots,s_p,1]\, . \label{Doddequivele}
 \eea

 \item{\framebox{$D$ odd:}}   Let $D=2p+1$, $p=1,2,\ldots$.  The little group is $SO(2p)$.  The only fields with propagating degrees of freedom are those with the following tableaux, corresponding to the permissible tableaux of $SO(2p)$, as detailed in appendix \ref{sodrepappendix},
 \bea && [0]\, ,  \ \ [s_1]\, , \  \ [s_1,s_2] \, ,\, \ldots\, , \ [s_1,\ldots,s_{p-1}] \, , \nn\\ 
&&  [s_1,\ldots,s_p]\, , \nn\\ 
&& [s_1,\ldots,s_{p-1},1,1] \, , \ \ldots \ ,   [s_1,s_2,1^{D-5}]\, , \ [s_1,1^{D-3}]\, ,\ \ [1^{D-1}] \, .\label{deDvenrepex2e}
\eea
Those in the first line are the minimal tensors and those in the third line are the non-minimal tensors.
We have the following duality equivalences among those in the first and third lines of \eqref{deDvenrepex2e},
 \bea && [0]\dualarrow  [1^{D-1}]\, ,\ \ [s_1]\dualarrow [s_1,1^{D-3}] \, ,\ \ [s_1,s_2]\dualarrow [s_1,s_2,1^{D-5}] \, ,\ \ldots\ ,  \nn\\
 &&[s_1,\ldots,s_{p-1}]  \dualarrow [s_1,\ldots,s_{p-1},1,1] \, .
 \eea

The $p$ row fields $[s_1,\ldots,s_p]$ are dual to themselves, 
\be \overset{\dualarrowcurved}{[s_1,\ldots,s_p]}\,,\ee
and can thus be split into chiral halves $[s_1,\ldots,s_p]_{\pm}$.
 \begin{itemize}
\item For $p$ even ($D=5,9,13,\ldots$) the self-duality relations are imaginary.   CPT flips the two representations, so both must be present in a theory.
\item For $p$ odd ($D=3,7,11,\ldots$) the self-duality relations are real.  CPT does not interchange the two representations and so each can appear separately in a theory, allowing for the existence of self-dual massive fields.   
\end{itemize}
 
 \end{itemize}
 
 The lowest few dimensions, which illustrate the general pattern above, are as follows:
 
 \begin{itemize}
 
 \item{\framebox{$D=2$:}}   The little group is trivial, and
 the only massive fields in two dimensions with propagating degrees of freedom are the scalar and Proca vector,
 \be [0]\, ,\ \ [1]\, .\ee
 They are dual to each other,
 \be [0]\dualarrow [1]\, .\ee
 This fact is familiar from the massless Schwinger model \cite{Schwinger:1962tp}, in which a 2D photon gets a mass by eating an electron-positron bound state, and the resulting dynamical field can be rewritten as a free massive scalar. 
 
 \item{\framebox{$D=3$:}}   The little group is $SO(2)$, the only massive fields with propagating degrees of freedom are the scalar, symmetric tensors and 2-form,
 \be [0]\, ,\ \ [s]\, ,\ \ [1,1]\, . \ee
The scalar and 2-form are dual to each other, and each carries the trivial representation of the little group,
\be  [0]\dualarrow [1,1]\,.\ee
 
The symmetric tensor fields $[s]$ are dual to themselves with real duality relations, 
\be \overset{\overset{{\rm Re}}{ \dualarrowcurved}}{[s]}\,,\ee
and can be split into chiral halves $[s]_\pm$.
These carry the helicity $\pm s$ representations of the $SO(2)$ little group.  Topologically massive electrodynamics for $s=1$ and topologically massive gravity for $s=2$ are well known theories that propagate only one of the two helicities \cite{Deser:1981wh}. 
 
\item{\framebox{$D=4$:}}   The little group is $SO(3)$, the massive fields with propagating degrees of freedom are 
 \be [0]\, ,\ \ [s]\, ,\ \ [s,1]\, ,\ \ [1,1,1]\, ,\ee
and the dualities between them are
 \be [0]\dualarrow [1,1,1]\, ,\ \ \  [s]\dualarrow [s,1]\, .\ee
The first of these is studied in \cite{Curtright:1980yj}, the case $s=1$ of second is the statement that a massive vector is dual to a massive 2-form, and the case $s=2$ is the statement that a massive Curtright field is dual to a massive graviton \cite{Curtright:1980yk,Curtright:1980yj,Morand:2012vx,Curtright:2019wxg}.

 \item{\framebox{$D=5$:}}   The little group is $SO(4)$, the massive fields with propagating degrees of freedom are,
\be [0]\, ,  \ \ [s_1]\, , \ [s_1,s_2] \, ,\ \ [s_1,1,1] \, , \ \ [1,1,1,1]\, . \ee
We have the dualities
\be [0] \dualarrow [1,1,1,1]\, ,\ \ \ [s_1]\dualarrow [s_1,1,1] \, .\ee
The two-row tensors $[s_1,s_2]$ dualize to themselves with imaginary coefficients,
\be \overset{\overset{{\rm Im}}{ \dualarrowcurved}}{[s_1,s_2]}\,,\ee
and can be split into chiral halves $[s_1,s_2]_\pm$.  Unlike in $D=3$, CPT interchanges the two representations that are carried by this field, so both must be present in a theory.

 \item{\framebox{$D=6$:}}   The little group is $SO(5)$, the massive fields with propagating degrees of freedom are,
 \be [0]\, ,\ \ [s_1]\, ,\ \ [s_1,s_2]\, ,\ \ [s_1,s_2,1]\,, \ \ [s_1,1,1,1]\,, \ \  [1,1,1,1,1]\, .\ee
We have the dualities
 \be [0]\dualarrow [1,1,1,1,1]\, ,\ \ \  [s_1]\dualarrow [s_1,1,1,1]\, ,\ \ \ [s_1,s_2]\dualarrow [s_1,s_2,1]\, .\ee

 \end{itemize}

\subsection{Massless flat space dualities\label{masslessdflatsection}}

 In the massless case, the representations are classified by the massless little group, which for spacetime dimension $D$ is $SO(D-2)$, and the general dualities can be understood in terms of these \cite{Bekaert:2002dt,deMedeiros:2002qpr}.  Two different fields are dual if their tableaux furnish equivalent representations of $SO(D-2)$, as reviewed in appendix \ref{sodrepappendix}.  From this we get the following dualities:

 \begin{itemize}
 
 \item{\framebox{$D$ odd:}}  Let $D=2p+3$, $p=0,1,2,\ldots$.  The little group is $SO(2p+1)$.  The only fields with propagating degrees of freedom are those with the following tableaux,
 \bea && [0]\, ,  \ \ [s_1]\, , \  \ [s_1,s_2] \, , \ldots\, , \ [s_1,\ldots,s_p] \, , \nn\\
&&  [s_1,\ldots,s_p,1] \, , \ \ldots \ ,  \ [s_1,s_2,1^{D-6}] , \ [s_1,1^{D-4}]\,  ,\  [1^{D-2}]\,. \label{smmodreeprDe} 
\eea
Those in the first row are the minimal tensors and those in the second row are the non-minimal tensors.
 We have the following duality equivalences,
 \be [0]\dualarrow  [1^{D-2}]\, ,\ \ [s_1]\dualarrow [s_1,1^{D-4}] \, ,\ \ [s_1,s_2]\dualarrow [s_1,s_2,1^{D-6}]\, ,\ldots,  [s_1,\ldots,s_p] \dualarrow [s_1,\ldots,s_p,1] . \label{Dmmoddequivele}\ee

 \item{\framebox{$D$ even:}}   Let $D=2p+2$, $p=0,1,2,\ldots$.  The little group is $SO(2p)$.  The only fields with propagating degrees of freedom are those with the following tableaux,
 \bea && [0]\, ,  \ \ [s_1]\, , \  \ [s_1,s_2] \, , \ldots, \ [s_1,\ldots,s_{p-1}]\,  , \nn\\ 
&&  [s_1,\ldots,s_p] \, , \nn\\ 
&& [s_1,\ldots,s_{p-1},1,1] \,, \ \ldots \ ,   [s_1,s_2,1^{D-6}] \, , \ [s_1,1^{D-4}] \, ,\ \ [1^{D-2}] \, .\label{demDvenrepex2e}
\eea
Those in the first line are the minimal tensors and those in the third line are the non-minimal tensors.  We have the following duality equivalences among those in the first and third lines,
 \bea && [0]\dualarrow  [1^{D-2}]\, ,\ \ [s_1]\dualarrow [s_1,1^{D-4}] \, ,\ \ [s_1,s_2]\dualarrow [s_1,s_2,1^{D-6}] \, ,\ \ldots\ , \nn\\
 && [s_1,\ldots,s_{p-1}] \dualarrow [s_1,\ldots,s_p,1,1] .
 \eea

The $p$ row fields $[s_1,\ldots,s_p]$ are dual to themselves, 
\be \overset{\dualarrowcurved}{[s_1,\ldots,s_p]}\,,\ee
and can thus be split into chiral halves $[s_1,\ldots,s_p]_{\pm}$.
 \begin{itemize}
\item For $p$ odd ($d=4,8,12,\ldots$) the duality relations are imaginary.   CPT flips the two representations, so both must be present in a theory.
\item For $p$ even ($d=2,6,10,\ldots$) the relations are real.  CPT does not interchange the two representations and so each can appear separately in a theory, allowing for the existence of self-dual massless fields.  
\end{itemize}

\end{itemize}

 The lowest few dimensions, which illustrate the general pattern above, are as follows:
 
 \begin{itemize}
 
 \item{\framebox{$D=2$:}}   The little group is trivial, and
 the only massless field in two dimensions with propagating degrees of freedom is the scalar
 \be [0]\, .\ee
It is dual to itself with real duality relation, 
\be \overset{\overset{{\rm Re}}{ \dualarrowcurved}}{[0]}\,,\ee
and the boson splits into chiral bosons $[0]_\pm$.  This is because the massless representation in $D=2$ splits into two irreducible pieces, namely the left and right moving momenta.

 \item{\framebox{$D=3$:}}   The little group is trivial and the only massless fields with propagating degrees of freedom are the scalar and vector,
 \be [0]\, ,\ \ [1]\, .\ee
They are dual to each other, 
\be  [0]\dualarrow [1]\,.\ee

\item{\framebox{$D=4$:}}   The little group is $SO(2)$ and the massless fields with propagating degrees of freedom are 
 \be [0]\, ,\ \ [s]\, ,\ \ [1,1]\, .\ee
The scalar and 2-form are dual
 \be [0]\dualarrow [1,1] \, .\ee
The massless spin-$s$ fields $[s]$ are self-dual with imaginary coefficients,
\be \overset{\overset{{\rm Im}}{ \dualarrowcurved}}{[s]}\,,\ee
so they split into chiral halves $[s]_\pm$.  Each half has 1 chiral degree of freedom, but because the self-duality relation is imaginary, both must be present in a theory. The case $s=1$ is classic electromagnetic duality, the cases $s\geq2$ are its extension to linearized gravity and massless higher spins.

 \item{\framebox{$D=5$:}}   The little group is $SO(3)$ and the massless fields with propagating degrees of freedom are,
\be [0]\, ,  \ \ [s]\ , \ [s,1]\,  \ \ [1,1,1] \, .\ee
We have the dualities
\be [0] \dualarrow [1,1,1]\, ,\ \ \ [s]\dualarrow [s,1] \, .\ee
The case $s=1$ is the statement that a massless 2-form is dual to a photon in five dimensions, and $s=2$ is the first instance of a dual graviton carried in a tensor representation different from that of the standard graviton.

 \item{\framebox{$D=6$:}}   The little group is $SO(4)$ and the massless fields with propagating degrees of freedom are,
\be [0]\, ,  \ \ [s_1]\, , \ [s_1,s_2] \, ,\ \ [s_1,1,1] \, , \ \ [1,1,1,1] \, .\ee
We have the dualities
\be [0] \dualarrow [1,1,1,1],\ \ \ [s_1]\dualarrow [s_1,1,1] \, .\ee
The two-row tensors $[s_1,s_2]$ dualize to themselves with real coefficients, 
\be \overset{\overset{{\rm Re}}{ \dualarrowcurved}}{[s_1,s_2]}\,,\ee
so they split into chiral halves $[s_1,s_2]_\pm$,
and we have the possibility of chiral fields propagating independently in a theory.  The best known example of this is the chiral 2-form, which appears as a degree of freedom in the tensor multiplet of $D=6$ superconformal field theories \cite{Heckman:2018jxk}.

\end{itemize}

\section{Tensor representations of orthogonal groups and their equivalences\label{sodrepappendix}}

Here we review the tensor representations of $SO(N)$, $O(N)$, and their equivalences, which are relevant for the discussions of duality.  

A tensor representation of $SO(N)$ is furnished by a tensor with indices in a fully traceless tableau with row lengths $[s_1,\ldots,s_p]$, where each index transforms by acting with the defining representation of $SO(N)$.  The tensor representations are honest finite dimensional group representations of $SO(N)$ (as apposed to the mere projective representations that the spinors furnish).  All such representations can be written as tensor representations for some tableau, but the map from tableau to representations is not bijective: depending on $N$, certain tensor representations can be trivial, equivalent to each other or reducible, as we detail in the following.

Following \cite{hamermesh1989group}, we call a tableau {\it permissible} if the sum of the lengths of its first two columns is $\leq N$.  A tableau is not permissible if and only if it has no free components and is trivial (generalizing the statement that a rank $r$ fully anti-symmetric tensor is trivial for $r>N$).  Thus we only need to consider the permissible tableau.  

We call two distinct tableau  {\it associated} if the length of the first column of the first tableau added to the length of the first column of the second tableau is $N$, and if the remaining columns of the two tableaux are identical.  Two associated tableau provide equivalent representations of $SO(N)$ 
 (this is the basis for all the flat space dualities). The equivalence is given by taking one of the tensors and contracting the $N$ dimensional epsilon tensor with the indices in the first column its tableau.  In a pair of associated tableaux, one always has more indices in its first column than the other.  We call the one with fewer indices in its first column the {\it minimal tensor} and the other the {\it non-minimal tensor}.  The non-minimal tensors can be considered redundant since they do not give distinct $SO(N)$ representations.

A tableau is {\it self-associate} if $N$ is even and the length of its first column is $N/2$.  Self-associate tensors break up into two inequivalent irreducible representations of $SO(N)$ given by forming self-dual and anti-self-dual combinations with respect to its first column.  These, plus the minimal tensors above, then give a one-to-one account of all $SO(N)$ irreducible representations (irreps).

The way this work in various dimensions, and how it extends to $O(N)$ by including parity\footnote{We take the parity transformation to be the transformation ${\rm diag}(1,\ldots,1,-1)\in O(N)$ that flips the $N$-th axis.  Other authors  sometimes call this the $R$ transformation, reserving $P$ for the transformation that flips all the axes.} is detailed as follows:

 \begin{itemize}
 
 \item{\framebox{$N$ odd:}}  Let $N=2p+1$, $p=0,1,2,3,\ldots$.
 The only permissible tableaux are 
\bea && [0]\, ,  \ \ [s_1]\, , \  \ [s_1,s_2]\,  , \ldots, \ [s_1,\ldots,s_p] \, , \nn\\
&&  [s_1,\ldots,s_p,1] \, , \ \ldots \ ,  \ [s_1,s_2,1^{N-4}]\,  , \ [s_1,1^{N-2}] \, ,\ \ [1^N]\,.\label{sodreepre} 
\eea
 
Among these, we have the following associated pairs,
 \be [0]\dualarrow  [1^N]\, ,\ \ [s_1]\dualarrow [s_1,1^{N-2}]\,  ,\ \ [s_1,s_2]\dualarrow [s_1,s_2,1^{N-4}] \, ,\ \ldots\ ,  [s_1,\ldots,s_p] \dualarrow [s_1,\ldots,s_p,1] \, . \label{doddequivele}\ee
 The tableaux in the first line of \eqref{sodreepre} are the {minimal tensors}, and those in the second line are the {non-minimal tensors}.
 The inequivalent irreps of $SO(2p+1)$ are in one-to-one correspondence with the minimal tensors, and they are all real irreps. 
 
 \underline{Parity}: The group $O(2p+1)$ is a direct product of $SO(2p+1)$ and the parity transformation $P$.  We thus get the reps of $O(2p+1)$ by lifting each tensor rep to 2 inequivalent reps where $P$ acts with a factor of $\pm 1$ in addition to the tensor transformation of the indices.   We call them tensors for $+1$ and pseudo-tensors for $-1$.  Due to the epsilon tensor, the sign of the $P$ action flips under the equivalences in \eqref{doddequivele}.
 
  \item{\framebox{$N$ even:}} Let $N=2p$, $p=1,2,3,\ldots$.
 The only permissible tableaux are 
\bea && [0]\, ,  \ \ [s_1]\, , \  \ [s_1,s_2]\, , \ldots, \ [s_1,\ldots,s_{p-1}]\, , \nn\\ 
&&  [s_1,\ldots,s_p] \, , \nn\\ 
&& [s_1,\ldots,s_{p-1},1,1]\,  , \ \ldots \ ,   [s_1,s_2,1^{N-4}] \, , \ [s_1,1^{N-2}] \, ,\ \ [1^N] \, .\label{devenrepex2e}
\eea
We have the following associated pairs among those in the first and third lines of \eqref{devenrepex2e},
 \bea && [0]\dualarrow  [1^N]\, ,\ \ [s_1]\dualarrow [s_1,1^{N-2}]\,  ,\ \ [s_1,s_2]\dualarrow [s_1,s_2,1^{N-4}] \, ,\ \ldots\ , \nn\\
 && [s_1,\ldots,s_{p-1}]  \dualarrow [s_1,\ldots,s_p,1,1] \, .
 \eea
The tableaux in the first line of \eqref{devenrepex2e} are the { minimal tensors}, and those in the third line are the {non-minimal tensors}; the minimal tensors all give distinct real irreps.

The tensors in the second line of \eqref{devenrepex2e} are self-associate, equivalent to themselves,
\be \overset{\dualarrowcurved}{[s_1,\ldots,s_p]}\,,\ee
and can be split into chiral halves,
\be [s_1,\ldots,s_p]_\pm\,,\ee
defined as follows (using the fact that the Hodge star satisfies $\ast^2=(-1)^p$ acting on $p$-forms in Euclidean signature):
\begin{itemize}
\item For $p$ even ($N=4,8,12,\ldots$) they are real (anti-)self-dual, 
\be T^\pm={1\over 2}\left(T\pm \ast T\right),\ \ \ \ast T^\pm=\pm T^\pm,\ \ \ T=T^++T^-,\ee
where the indices on $T\in  [s_1,\ldots,s_p]$ are suppressed and the star operator acts only on the indices of the first column of the tableau.
 Each chiral half is a real irrep of $SO(N)$.  Together with the minimal tensors above, they give all irreps of $SO(N)$, which are thus all real.    
\item For $p$ odd ($N=2,6,10,\ldots$) they are imaginary (anti-)self-dual, 
\be T^\pm={1\over 2}\left( T\mp i\ast T\right)\,, \ \ \ast T^\pm=\pm i \,T^\pm\,,\ \ \ T=T^++T^-\,. \ee
Each part is a complex irrep, and the two parts are conjugate to each other.   The direct sum of the two, i.e. the original tensor $T$,  is a real irrep.  Together with the minimal tensors above, they give all irreps of $SO(N)$,
\end{itemize}

 \underline{Parity}: The group $O(2p)$ is a not a direct product of $SO(2p)$ and the parity operator $P$; it is only a semi-direct product.  
 
\begin{itemize}
\item For $p$ even ($N=4,8,12,\ldots$) all the $SO(N)$ reps are real and are lifted to two different reps of $O(N)$ where $P$ acts with a factor of $\pm 1$ in addition to the tensor transformation of the indices.   In particular, $P$ does not interchange the two reps $[s_1,\ldots,s_p]_\pm$, so that each individually remains irreducible and lifts to an irrep of $O(N)$.
\item For $p$ odd ($N=2,6,10,\ldots$) all the reps except $[s_1,\ldots,s_p]_\pm$ are real and are lifted to two different reps of $O(N)$ where $P$ acts with a factor of $\pm 1$ in addition to the tensor transformation of the indices.  The complex reps $[s_1,s_2,\ldots,s_p]_\pm$ are joined together into one irrep of $O(N)$, with $P$ interchanging the two in addition to its action on the indices; this is a real irrep of $O(2p)$.
\end{itemize} 
 
\end{itemize}

Here we detail the lower dimensional cases, for ease of reference and to illustrate the general pattern above:
 
 \begin{itemize}
\item{\framebox{$N=1$:}}  This case is degenerate since the group $SO(1)$ is trivial.  The only permissible tableaux for $N=1$ are the scalar and vector, 
\be [0]\, ,\ \ [1]\, ,\ee
and they are dual to each other by simply equating the single component of the vector with the scalar (this can be considered the action of the one-dimensional epsilon symbol),
\be [0] \dualarrow [1]\, .\ee
They both carry the trivial representation.  The non-minimal tensor is $[1]$ and the minimal tensor is $[0]$.

 \underline{Parity}: The group $O(1)$ is isomorphic to ${\mathbb Z}_2$, with parity $P$ the non-trivial element. The inequivalent irreps are scalars and pseudo-scalars.  Scalars are equivalent to pseudo-vectors, and pseudo-scalars are equivalent to vectors.

\item{\framebox{$N=2$:}}   The only permissible tableaux are the scalar, fully symmetric traceless tensors, and 2-form,
\be [0]\, , \ [s] \, , \ [1,1]\, .\ee
The scalar is associated to $[1,1]$ through the two dimensional epsilon symbol,
\be [0] \dualarrow [1,1]\, ,\ee
and both carry the trivial representation of $SO(2)$.  $[0]$ is the minimal tensor and $[1,1]$ is the non-minimal tensor.
The symmetric tensors are self-associate and dualize to themselves, 
\be \overset{\overset{{\rm Im}}{ \dualarrowcurved}}{[s]}\,.\ee
They spit into imaginary self-dual and anti-self-dual parts,
\be [s]_\pm  \,.  \ee
These are the helicity $\pm s$ irreps of $SO(2)$, they are complex one-dimensional irreps which are conjugate to each other.
 
  \underline{Parity}: The group $O(2)$ is a semi-direct product of $SO(2)$ and $P$, not a direct product.  The scalar rep lifts to scalar and pseudo-scalar representations of $O(2)$, equivalent to pseudo 2-form and 2-form representations respectively.  For each $s$ the two helicity reps $[s]_\pm$ are mapped into each other under $P$.

\item{\framebox{$N=3$:}}    The only permissible tableaux are 
\be [0]\, ,  \ \ [s]\, , \ \ [s,1] \, ,\ \ [1,1,1]\, .\ee
Through the $N=3$ epsilon symbol we have the associated pairs
\be [0] \dualarrow [1,1,1]\, ,\ \ \ [s]\dualarrow [s,1] \, .\label{d3equivalencesee}\ee
$[0]$ is the minimal tensor carrying the trivial representation of $SO(3)$ and $[1,1,1]$ is the non-minimal tensor.   $[s]$ is the minimal tensor carrying the spin $s$ representation of $SO(3)$ and $[s,1]$ is the non-minimal tensor.  All these representations are real.
 
\underline{Parity}: The group $O(3)$ is a direct product of $O(3)$ and $P$, $P$ acts with $\pm 1$ giving tensors and pseudo-tensors.  Tensors are equivalent to pseudo-tensors and vice-versa under the equivalences \eqref{d3equivalencesee}.
 
\item{\framebox{$N=4$:}}     The only permissible tableaux are 
\be [0]\, ,  \ \ [s_1]\ , \ \ [s_1,s_2]\,  ,\ \ [s_1,1,1] \, , \ \ [1,1,1,1] \, .\ee
Through the $N=4$ epsilon symbol, we have the associated pairs
\be [0] \dualarrow [1,1,1,1]\, ,\ \ \ [s_1]\dualarrow [s_1,1,1] \, .\ee
The minimal tensors are those on the left hand side of each pair and the non-minimal tensors are those on the right hand side.
The two-row tensors dualize to themselves, 
\be \overset{\overset{{\rm Re}}{ \dualarrowcurved}}{[s_1,s_2]}\,,\ee
and can be spit into real self-dual and anti-self-dual parts,
\be [s_1,s_2]_\pm\,.\ee
These parts are each real irreps of $SO(4)$, and so all the irreps of $SO(4)$ are real.

  \underline{Parity}: The group $O(4)$ is semi-direct product of $SO(4)$ and $P$, not a direct product.  All the irreps lift to tensor and pseudo-tensor irreps of $O(4)$ where $P$ acts with $\pm 1$.  In particular, the $[s_1,s_2]_\pm$ are not mapped into each other, and remain irreducible under $O(4)$.
    
\item{\framebox{$N=5$:}}     The only permissible tableaux are 
\be [0],  \ \ [s]\, , \ \ [s_1,s_2] \, , \, \ \ [s_1,s_2,1] \,,  \  \ [s,1,1,1] \, ,\ \ \ [1,1,1,1,1] \,.\ee
Through the $N=5$ epsilon symbol we have the associated pairs
\be [0] \dualarrow [1,1,1,1,1]\, ,\ \ \ [s]\dualarrow [s,1,1,1] \, , \ \  \ \ [s_1,s_2]\dualarrow  [s_1,s_2,1]\, . \label{d5equivalencesee} \ee
The minimal tensors are those on the left hand side of each pair and the non-minimal tensors are those on the right hand side.
All these representations are real.

\underline{Parity}: The group $O(5)$ is a direct product of $O(5)$ and $P$, all the representations lift to tensor and pseudo-tensor representations of $O(5)$ where $P$ acts with $\pm 1$.

\item{\framebox{$N=6$:}}     The only permissible tableaux are 
\be [0]\, ,  \ \ [s]\,,\ \ [s_1,s_2] \, , \ \   \ [s_1,s_2,s_3] \, , \ \ [s_1,s_2,1,1] \, ,\ \ [s,1,1,1,1] \, ,\ \ \ [1,1,1,1,1,1] \, . \ee
Through the $N=6$ epsilon symbol we have the associated pairs
\be [0] \dualarrow [1,1,1,1,1,1]\, , \ [s]\dualarrow [s,1,1,1,1] \,,  \ [s_1,s_2]\dualarrow  [s_1,s_2,1,1] \, . \ee
The minimal tensors are those on the left hand side of each pair and the non-minimal tensors are those on the right hand side.  All these representations are real.

The three-row tensors are self-dual
\be \overset{\overset{{\rm Im}}{ \dualarrowcurved}}{[s_1,s_2,s_3]}\,,\ee
and can be spit into imaginary self-dual and anti-self-dual parts,
\be [s_1,s_2,s_3]_\pm\,.\ee
These parts are each complex irreps of $SO(6)$.  

  \underline{Parity}: The group $O(6)$ is a semi-direct product of $SO(6)$ and $P$, not a direct product.  All the reps except the 3-row reps lift to tensor and pseudo-tensor irreps of $O(6)$.  The 3-row chiral reps $[s_1,s_2,s_3]_\pm$ are mapped into each other under $P$, and taken together give a real rep of $O(6)$.

\end{itemize}

\newpage

\bibliographystyle{utphys}
\addcontentsline{toc}{section}{References}
\bibliography{PMshiftduality-arxiv-v2}

\end{document}